\documentclass[12pt]{article}
\usepackage[utf8]{inputenc}
\usepackage[margin=3cm]{geometry}
\usepackage{amsmath, amsfonts, amssymb}
\usepackage{color, hyperref}
\usepackage{graphicx, epstopdf}
\usepackage[absolute]{textpos}
\usepackage{cite}
\usepackage{tabularx}
\usepackage{subcaption}
\usepackage{appendix}
\usepackage{hyperref}
\usepackage{ragged2e}
\usepackage[capitalise]{cleveref}
\usepackage[labelfont=bf, labelsep=period, font=small]{caption}

\usepackage{contour}

\definecolor{myred}{cmyk}{0,1,1,0.55}
\definecolor{mygreen}{rgb}{0.27, 0.64, 0.48}
\definecolor{mygray}{gray}{.95}

\begin{document}

\rightline{DESY-24-193}
\vspace{10mm}

\begin{center}
{\bf\Large Non-singular solutions to the Boltzmann equation with a fluid Ansatz} \\
[5mm]
\renewcommand*{\thefootnote}{\fnsymbol{footnote}}
Gl\'auber C. Dorsch$^{a}$\footnote{\href{mailto:glauber@fisica.ufmg.br}{glauber@fisica.ufmg.br}} 
Thomas Konstandin$^{b}$
\footnote{\href{mailto:thomas.konstandin@desy.de}{thomas.konstandin@desy.de}},
Enrico Perboni$^{b}$
\footnote{\href{mailto:enrico.perboni@desy.de}{enrico.perboni@desy.de}}
and Daniel A. Pinto$^{a}$\footnote{\href{mailto:dwavea@ufmg.br}{dwavea@ufmg.br}} 
\\
\vspace{2mm}
$^{a}$\,{\it Universidade Federal de Minas Gerais, 31270-901, Belo Horizonte, MG, Brazil.}
\\
\vspace{1mm}
$^{b}$\,{\it Deutsches Elektronen-Synchrotron DESY, Notkestr.~85, 22607 Hamburg, Germany.}
\date{}
\end{center}

\begin{center}
\vspace{1cm}
    {\bf Abstract}
    
    \vspace{5 mm}
    \justifying{
    Cosmological phase transitions 
    can give rise to intriguing phenomena, such as baryogenesis or a stochastic  gravitational wave background, due to nucleation and percolation of vacuum bubbles in the primordial plasma. A key parameter for predicting these relics is the bubble wall velocity, whose computation relies on 
    solving the Boltzmann equations of the various species
    along the bubble profile. Recently it has been shown that an unphysical singularity emerges if one assumes these local quantities to be described as small fluctuations over a constant equilibrium background. 
    In this work we solve this issue by including the spatial dependence of the 
    background into the fluid \emph{Ansatz}. This leads to a modification of the Boltzmann equation, and all terms that would give rise to a singularity now vanish.
    We recalculate the different contributions to the counter-pressure of the plasma on the expanding wall, and discuss their relative importance. The Standard Model with a low cutoff is chosen as benchmark model and the results are shown for different values of the cutoff scale $\Lambda$. In this setup, deflagration solutions are found for almost all the values of $\Lambda$ considered, while detonations are found only for some restricted corner of the parameter space.  }
\end{center}

\renewcommand*{\thefootnote}{\arabic{footnote}}
\setcounter{footnote}{0}

\newpage
\hrule
\tableofcontents

\vspace{3em}
\hrule
\vspace{2em}

\newpage
\section{Introduction} 
The discovery of gravitational waves (GWs) by LIGO, VIRGO and Kagra ~\cite{LIGOScientific:2016aoc, LIGOScientific:2018mvr, LIGOScientific:2021usb, LIGOScientific:2021djp} and the strong evidence for a stochastic GW background recently observed by several pulsar timing arrays, like NANOGrav \cite{NANOGrav:2023gor}, EPTA and InPTA \cite{EPTA:2023fyk}, have urged the community to understand the different mechanisms that could give rise to these signals during the evolution of the Universe. First-order cosmological phase transitions play a prominent role among them, especially because of the implication they would have for particle physics. In fact, when a phase transition is first order, it proceeds via bubble nucleation, expansion and percolation, producing sound waves and turbulence in the plasma, which (together with the kinetic energy of the Higgs field) could source stochastic gravitational waves. The same process could also give origin to other relics, such as a dark matter abundance or a matter-antimatter asymmetry, so the detection of such a primordial radiation would give us also some important information about the underlying physics model, becoming a complementary source of information to those obtainable from collider experiments~\cite{Caprini_2020,Arcadi_2024}.

Crucially, the abundance of these relics depends, among other few parameters, on the expansion velocity of these bubbles, $v_w$. Determining whether they will reach a terminal wall velocity or whether they will keep accelerating up to the so-called runaway behavior (with $v_w \sim 1$~\cite{Bodeker:2009qy}) depends on a precise modeling of the counter-pressure acting against the expanding wall. 
In recent years, much progress has been made in this direction. In the limit of runaway walls {the so-called ballistic approximation} can determine the backreaction against it~\cite{Bodeker:2009qy, Hoche:2020ysm, Azatov:2020ufh, Ai:2024shx, Long:2024sqg}, because the wall advances so fast that the plasma has not received any information about its arrival yet, and {the distribution is only modified by kinematic considerations}. But for non-runaway behaviour the question of how much the out-of-equilibrium effects impact the calculation of $v_w$ is more involved. In this case one needs to compute the out-of-equilibrium distribution functions, which involves solving the integro-differential Boltzmann equation. 
This is a numerically expensive task, and to make it more efficient, it is convenient (though not mandatory~\cite{DeCurtis:2022hlx, DeCurtis:2023hil, DeCurtis:2024hvh}) to choose a particular \textit{Ansatz} for the shape of the non-equilibrium fluctuations. If one is able to solve the collision integrals in terms of the parameters in this \emph{Ansatz}, one can convert the integro-differential equation into a more manageable set of ordinary differential equations, which can be solved numerically in a matter of seconds on a desktop computer. 

Since the 1990s one popular choice has been the so-called \textit{fluid Ansatz}~\cite{Moore:1995si, Konstandin:2014zta, Dorsch:2021ubz, Dorsch_2022, dorsch_bubble_2023}, where the out-of-equilibrium distribution takes the same form as the Bose-Einstein/Fermi-Dirac functions but with a slightly modified argument to account for deviations from equilibrium, which can then be expanded in powers of momenta, i.e.
\begin{equation}
    f_i(x,p) = \dfrac{1}{e^{\beta(u^\mu p_\mu - \delta_p)}\pm 1}, \quad\text{with}\quad \delta_p = w^{(0)} + w^{(1)}_\mu p^\mu + w^{(2)}_{\mu\nu} p^\mu p^\nu+\ldots,
    \label{eq:fluid}
\end{equation}
with $u_\mu$ the plasma four-velocity. This choice has the advantage that fluctuations $w^{(i)}$ have a clear physical interpretation in terms of the local chemical potential, temperature, fluid velocity and (for $\mathcal{O}(p^2)$) dissipative effects~\cite{DeGroot:1980dk}. {This approach is up to this point rather generic and also other choices for the deviations can be found in the literature.}
In any case, in order to make the Boltzmann equations more tractable, one further assumes that the fluctuations are small compared to the background (i.e. equilibrium) values, and performs a \emph{linearization} of the equations. It turns out that, if the background is not modelled correctly, this procedure leads to a singularity in the behaviour of the plasma friction when the wall approaches the speed of sound. This is not a problem of this specific \emph{Ansatz}, but rather of the linearization procedure and of the inadequate choice of the background. 
It happens because of an interplay between energy-momentum conservation and the inclusion of fluctuations representing collective shifts in temperature and fluid velocity, both being \textit{Ansatz}-independent conditions.

An important contribution towards solving this singularity issue was made in ref.~\cite{laurent_first_2022}. Recognizing that the problem arises from the background fluctuations, the authors first use hydrodynamical considerations to solve the energy-momentum tensor conservation of the fluid and find the velocity and temperature profiles across the phase transition front. The non-equilibrium fluctuations are then modelled around these profiles (rather than assuming a constant background), and the Boltzmann equations are solved for the top quark using a \textit{spectral Ansatz}, expanding in terms of Chebyshev polynomials. 
The disadvantage, in this case, lies in losing the physical interpretation provided by the fluid \emph{Ansatz}. {More concretely, there are no basis elements that 
represent a local change of fluid velocity or temperature. Besides, it also requires a fully numerical computation of the collision terms. We will comment on these and other differences again in the conclusion section.}

In this work we follow a double-step procedure similar to ref. \cite{laurent_first_2022}, but stick to the \textit{fluid Ansatz} and solve the Boltzmann setup not only for the top quark, but also taking into account the heavy gauge bosons ($W^\pm$ and $Z^0$) and the fluctuations of the light elements. The former can give important corrections to the final value of $v_w$ while the latter need to be studied because they were the ones leading to the singularity in the {previous} approach, even if we expect them to give a subleading contribution to the friction.
We find that, using this framework, the singularity in the friction disappears, while a discontinuity remains at the Jouguet velocity, as expected, corresponding to crossing from the regime of supersonic deflagrations (hybrids) to detonations. 

Because the background temperature and velocity profiles now depend on their distance to the wall, they will also give a contribution to the fluid pressure (which generally depends on $\partial_z T$). This corresponds to a pressure stemming purely from equilibrium considerations, and is well known in the literature~\cite{Konstandin:2010dm}. Lately there have been some claims that this term outweighs the  friction term stemming from non-equilibrium fluctuations around the background, so that purely equilibrium considerations would suffice for an accurate estimate of the wall velocity~\cite{Ai:2021kak,laurent_first_2022,Ai:2023see}. Our results do not fully corroborate this claim. If we considered only the equilibrium contribution then all detonations would necessarily be runaways~\cite{Lewicki:2021pgr}, {while including the non-equilibrium pressure can lead to stationary solutions}. However, for the specific toy-model we study here (with an effective $\phi^6/\Lambda^2$ term in the scalar potential) this happens only for tuned parameters in the range $\Lambda \in [625 - 635]$~GeV. Moreover, for every $\Lambda$ one also finds a deflagration solution, which is most likely the stable solution and detonations are never actually realized {in nature}. These results are consistent with other recent works~\cite{Cline:2021iff, Lewicki:2021pgr}.

The paper is organized as follows. In \cref{sec. lin.BEs} we review the formalism that leads to the linearized system of Boltzmann equations and show that the singularity stems from the fact that the equations corresponding to energy-momentum conservation
of the fluid contain a source. In \cref{sec. new_way_Boltz} we present our new method for dealing with the Boltzmann equation, now with a spatial dependence of the background fluctuations. These are computed by absorbing the sources of the total fluid energy-momentum tensor in the Boltzmann equation into the background temperature and fluid velocity, leading to a non-singular behaviour. In \cref{sec. hydrodyn.} we review the hydrodynamics of the plasma-wall system, focusing especially on the temperature and velocity behavior around the wall in the three possible cases of deflagrations, detonations and hybrids. 
In \cref{sec:: friction} we discuss the Higgs equation of motion in the presence of a plasma out of equilibrium. This equation contains terms that are interpreted as counter-pressures of the fluid on the expanding wall. We discuss the equilibrium and non-equilibrium contributions to this pressure. In \cref{sec. results and discussion} we discuss our findings and some of the approximation used throughout the paper. Finally, our conclusions are presented in \cref{sec. conclusions}. 

\section{Linearized Boltzmann setup}
\label{sec. lin.BEs}

Let us first recapitulate the linearized approach to solving the Boltzmann equations used in references~\cite{Konstandin:2014zta, Dorsch_2022, dorsch_bubble_2023}, and see how it leads to the singularity across the speed of sound.

Recall that we are working in a scenario where the Universe undergoes a first-order phase transition, and vacuum bubbles nucleate and expand in the primordial plasma. As the bubble passes by, it drives the plasma out of equilibrium. If the bubble wall width is much larger than the inverse momenta of the incident particles from the plasma, $L_w \gg 1/p \sim 1/T$, a WKB approximation can be applied and it can be shown~\cite{Konstandin:2013caa} that the Kadanoff-Baym equation for the two-point function (encoding information on the particle distribution functions $f_i\left(x^\mu, p^\mu\right)$) reduces to
the semi-classical Boltzmann equation 
\begin{equation} 
    p^\mu \partial_\mu f_i\left(x^\mu, p^\mu\right)+m F^\mu \partial_{p^\mu} f_i\left(x^\mu, p^\mu\right)+\mathcal{C}\left[f_j\right]=0,
    \label{eq. BEq}
\end{equation}
where $p^\mu$ is the on-shell particle four-momentum, $F^\mu = \partial^\mu m$ is the semi-classical force driving the particles out of equilibrium, and $\mathcal{C}$ are the collision terms which tend to bring the system back to equilibrium. Our task is to find a solution to this equation, since deviations from equilibrium will result in a pressure on the bubble wall which affects the final velocity $v_w$. 

Without loss of generality, one can always decompose the distribution functions as a sum of an equilibrium and a non-equilibrium part,
\begin{equation}
    f_i\left(x^\mu, p^\mu\right) =  \frac{1}{e^{\beta(p^{\mu } u_{\mu} - \delta_p - \delta_{p,\text{light}})}}  = f_i^{\text{eq}} \left( \beta p^{\nu} u_{\nu} \right) + \delta f_i\left(x^\mu, p^\mu\right),
\end{equation}
with $\beta^{-1} = T$ being the temperature of the plasma and $u_\mu$ its four-velocity relative to the observer. Note (and this is crucial) that we have split the fluctuations as $\delta_p + \delta_{p,\text{light}}$: the latter are fluctuations of the light species relative to the background, and the former are fluctuations of the heavy species relative to the light ones and the background. {Here, the \emph{background is assumed to be constant}, unlike the approach that we will follow in later sections.} 

Now, in order to solve the Boltzmann equation for the fluctuations $\delta f_i$, one first needs to find an efficient way to compute the collision terms\footnote{See refs.~\cite{DeCurtis:2022hlx,DeCurtis:2023hil, DeCurtis:2024hvh} for a fully numerical method to compute the collision integrals. The idea is essentially to solve eq.~\eqref{eq. BEq} on a lattice in momentum space, exploiting symmetries of the collision terms that allow one to reduce the corresponding 9-dimensional collision integral down to two dimensions. However, even after these simplifications, finding a solution on this lattice is numerically very expensive, taking about one hour to find one solution using a desktop computer according to the authors of this approach. To find $v_w$ one usually has to iteratively solve the Boltzmann equation, as we shall discuss below, which then leads to even larger costs in computational time. The problem is then further aggravated if one wants to find $v_w$ while scanning over some parameter space. A semi-analytic approach that yields reliable results would be better suited for this task.}, and then rewrite the Boltzmann equation in a more manageable form in terms of the fluctuations $\delta_p$. For this purpose we assume that the fluctuations are small relative to the background, and we truncate the momentum expansion of $\delta_p$ (cf. eq.~\eqref{eq:fluid}) down to first order terms only, so that the fluctuations are parametrized as
\begin{equation}
    \delta_p = \delta \mu + p^\mu ( \delta u_\mu - u_\mu \delta T / T ), 
    \label{eq:truncation}
\end{equation}
and similarly for $\delta_{p,\text{light}}$. In this way $\delta \mu$ corresponds to the particle's chemical potential, $\delta T / T$ tracks fluctuations in the fluid temperature and $\delta u_\mu$ (which satisfies $u^\mu \delta u_\mu = 0$) are the velocity fluctuations. 

One assumes furthermore that the bubble has expanded enough so that the effects of the wall curvature are negligible, and the problem becomes one-dimensional. In the rest frame of the bubble wall, with the $z$-axis orthogonal to the wall and oriented towards the broken phase, one then has $u^\mu = \gamma_w (1, v_w)$, $\delta u^\mu \equiv \delta v\, \bar{u}_\mu = \delta v \gamma_w(v_w,1)$ and $m^2(x) \to m^2(z)$. Then, since we have three fluctuations, one takes three moments of the Boltzmann equation, namely multiplying it by $1$, $p^\mu u_\mu$, $p^\mu \bar u_\mu$ (where $\bar u_\mu$ is the four velocity orthogonal to $u_\mu$) and integrating over $p$. 

The assumption that the fluctuations are small relative to the background allows one to simplify the collision terms to a linearized form $\Gamma\cdot q$ and reduce the Boltzmann equation to a set of ordinary differential equations for $q=(\delta \mu, -\delta T/T, \delta v)^T$~\cite{Moore:1995si, Konstandin:2014zta, Dorsch_2022}. The matrix $\Gamma$ encapsulates the collision rates coming from integrals in the collision term $\mathcal{C}[f_i]$, which can be calculated numerically, although simple analytical results are also viable if one assumes a leading-log approximation~\cite{Moore:1995si, Dorsch:2021ubz}. 

The source terms arise from the semi-classical force term $F^\mu = \partial^\mu m$ in the Boltzmann equation, as well as from the kinetic term (i.e. the Liouville operator) due to the $z$-dependence in the momentum $p^\mu$ (recall $p^0$ depends on the particle's mass). Importantly, these terms are absent for the light species by construction, since their masses are not significantly changed during the passage of the bubble. The fluctuations of the light species are only sourced indirectly due to collision terms with the heavy particles. Moreover, the chemical potential of the light species vanish identically since their particle numbers equilibrate quickly, so the light elements can be described only through velocity and temperature fluctuations. This means that the corresponding moments of the Boltzmann equation for these
species involve a multiplication by $p^\mu u_\mu$ or $p^\mu \bar u_\mu$, with a single power of $p^\mu$. Upon integration, the resulting equation corresponds to the macroscopic energy-momentum conservation condition. Because of this, when summing over the equations for \emph{all} the fluctuations (heavy and light species alike), the sum over the collision terms must vanish. This means that the collision matrix of the light species with tops and weak gauge bosons (collectively denoted as $W$) satisfy~\cite{Dorsch_2022}
\begin{equation}
    \Gamma_{\text{light},t} = -N_t \Gamma_t
    \quad\text{and}\quad
    \Gamma_{\text{light},W} = -N_W\Gamma_W,
    \label{eq:lightcollision}
\end{equation}
with $N_{t,W}$ the respective number of degrees of freedom, and $\Gamma_{t,W}$ the collision matrix entering the Boltzmann equation for the heavy fluctuations. We will soon see that these relations play a major role leading to the singularity problem in this linearized procedure. 

With these considerations, the Boltzmann equation~\eqref{eq. BEq} for the case of a Standard Model (SM) like plasma becomes\cite{Konstandin:2014zta, Dorsch_2022}
\begin{equation} \label{eq. Boltz. eq heavy guys OLD}
\begin{aligned}
A_W \cdot\left(q_W+q_{\mathrm{light}}\right)^{\prime}+\Gamma_W \cdot q_W & =S_W, \\
A_t \cdot\left(q_t+q_{\mathrm{light}}\right)^{\prime}+\Gamma_t \cdot q_t & =S_t, \\
A_{\mathrm{light}} \cdot q^\prime_{\mathrm{light}}+\Gamma_{\mathrm{light}, W} \cdot q_W+\Gamma_{\mathrm{light}, t} \cdot q_t & =0,
\end{aligned}
\end{equation}
where prime denotes the derivative with respect to the $z$ coordinate. The kinetic matrices $A_W$, $A_t$ and $A_{\text{light}}$ take the form 
\begin{equation} \label{eq:: A_matrix}
A_i=\frac{\gamma_w}{2 \pi^2} \left(\begin{array}{lll}
v_w c_2 & v_w c_3 & c_3 / 3 \\
v_w c_3 & v_w c_4 & c_4 / 3 \\
c_3 / 3 & c_4 / 3 & v_w c_4 / 3
\end{array}\right), \quad A_{\mathrm{light}}=\gamma_w \frac{c_4^{\mathrm{light}}}{2 \pi^2}\left(\begin{array}{cc}
v_w & 1 / 3 \\
1 / 3 & v_w / 3
\end{array}\right),
\end{equation}
with coefficients $c_n$ differing for bosons and fermions and defined as
\begin{equation}
    c_1^f = \log2
    \quad\text{and}\quad 
    c_1^b=\log(2T/m)
\end{equation}
and for $n\geq 2$ as
\begin{eqnarray}
c_n^b \equiv \frac{1}{T^{n+1}}\int dp \, p^n  f^\text{BE}_p (1+ f^\text{BE}_p) &=& n!\,\zeta_n\,,\\
c_n^f \equiv \frac{1}{T^{n+1}}\int dp \, p^n  f^\text{FD}_p (1- f^\text{FD}_p) &=& \left(1-\frac{1}{2^{n-1}}\right) n!\,\zeta_n\,,
\label{eq:p_ints}
\end{eqnarray}
and $f^{\text{BE},\, \text{FD}}$  the Bose-Einstein and Fermi-Dirac equilibrium distributions. The kinetic matrix for the light species is the bottom right $2 \times 2$ block of the $A_i$ matrix, summed over their fermionic and bosonic degrees of freedom, such that $c_4^{\text{light}} = 78 c_4^f + 19 c_4^b$. 
The matrices $\Gamma_X$ and $\Gamma_{\text{light},X}$ are respectively $3 \times 3$ and $2 \times 3$ and represent the collision terms. The source terms are given by 
\begin{equation} \label{eq. old_sources}
S=\gamma_w v_w \frac{m' m}{ 2 \pi^2 T^2}\left(\begin{array}{c}
c_1 \\
c_2 \\
0
\end{array}\right).
\end{equation}
It is important to notice that, in the classical setup, the presence or not of a source term depends on the mass gain during the phase transition. This means that (a) no source term is present for the light species and (b) no contribution to the sources come from the term $p^\mu \partial_\mu f_i^{bg}$ because the equilibrium distribution functions are considered space independent~\footnote{In fact there is a term from the Liouville operator, which however cancels against a force term. This cancellation ensures that the remaining force term in (\ref{eq. old_sources}) vanishes for zero wall velocity. }.

\subsection*{The origin of the singularity} 

It turns out that the solution of the equations for the light species will be singular at the speed of sound, leading to a singularity in the friction as well. As pointed out in\cite{Dorsch_2022}, this singularity is due to an interplay between energy momentum conservation and a zero eigenvalue in the kinetic matrices at $v_w = c_s$. To see how this comes about, recall from the discussion above that we took two moments of the Boltzmann equation which correspond to energy-momentum conservation, and from this one obtains a relation between collision terms as given in~\cref{eq:lightcollision}. 
This means that there are two linear combinations of the equations (\ref{eq. Boltz. eq heavy guys OLD}) that give us a linearized version of the conservation of these quantities. In other words, there are two vectors $\chi$ such that $\chi \cdot \Gamma = 0$ and 
\begin{equation} \label{eq. linearized_en_mom_cons}
    \chi \cdot A \cdot q' = \chi \cdot S \, .
\end{equation} 
Integrating these over $z$ gives\begin{equation} \label{eq. linear-en-mom}
    \chi \cdot A \cdot \Delta q = \chi \int S \, dz.
\end{equation}
The source term parameterizes the difference in the equation of state between the two phases due to the mean field term, and can be integrated to obtain $p_+ - p_- \propto m^2 T^2 \propto \int S dz$. At the same time, almost all the fluctuations are damped far away from the wall, due to the collision terms. The only exceptions are the ones that parameterize a collective shift in the local equilibrium, $\delta T_{\text{light}}$ and $\delta v_{\text{light}}$. This means that \cref{eq. linear-en-mom} becomes a constraint for these two fluctuations, which can be solved for the their change across the bubble wall, giving \cite{Dorsch_2022}
\begin{equation} \label{eq. constraint-vwall}
    \delta v_{\text{light}} = \frac{\sum_i N_i m_i^2 c_2^i}{2 T^2 \sum c_4} \frac{3 v_w}{1 - 3 v_w^2}, 
\end{equation}
where the sum over $i$ runs over the heavy particles (tops, $W^\pm$ and $Z^0$) while the sum over $c_4$ runs over all the degrees of freedom. The singularity at $v_w=c_s=1/\sqrt{3}$ then becomes explicit.

These two relations show that the singularity is related to a conceptual issue: the equations that should represent the linearized version of the energy-momentum tensor conservation are \textit{sourced} by the mean field term in the potential. This does not depend on the fact that we truncated the momentum expansion in~\cref{eq:truncation} at three fluctuations: even considering a \textit{generalized fluid Ansatz}~\cite{Dorsch:2021ubz, Dorsch_2022, dorsch_bubble_2023} would not solve the issue. Instead, it relies on the fact that we absorbed in the fluctuations of the light species the collective changes in the local equilibrium quantities. If these background fluctuations are large, this will lead to a breakdown of the linearization procedure.

\subsection*{A hint from hydrodynamics} 

To have an idea of what is going wrong with the usual linearization procedure presented in \cite{Moore:1995si, Konstandin:2014zta}, we can look at what happens to the plasma-wall system on length scales much larger than the longest damping scale in the process. The separation of scales is easily achieved, since the wall thickness is many orders smaller than the bubble size. The hydrodynamics of the \textit{plasma-wall} system can be studied through the energy-momentum tensor of the system \cite{Espinosa:2010hh}. The energy-momentum tensor of the Higgs field $\phi$ is given by
\begin{equation} \label{eq:: higgs en_mom_tensor}
    T_{\mu \nu}^\phi = \partial_\mu \phi \partial_\nu \phi - g_{\mu \nu} \left( \frac{1}{2} \partial_\sigma \phi \partial^\sigma \phi - V_0(\phi) \right),
\end{equation}
where $V_0(\phi)$ is the zero temperature vacuum potential. The energy-momentum tensor of the plasma is given by 
\begin{equation}
    T_{\mu \nu}^{\text{plasma}} = \sum_i \int \frac{d^3 p}{(2 \pi)^3E_i} p_\mu p_\nu f_i(p,x),
\end{equation}
where the sum is carried out over all the species in the plasma, $f_i(p,x)$ are the particle distribution functions and $p^\mu$ is again the on-shell particle four-momentum, related to the energy $E$ through $p^0 = E =\sqrt{\vec{p}^2+m^2}$. 
If the plasma is in local equilibrium, this can be parameterized as 
\begin{equation} \label{eq. equil.energy-mom tensor}
    T_{\mu\nu}^{\text{plasma}} = \omega u_\mu u_\nu - P g_{\mu \nu},
\end{equation}
where $P$ and $\omega\equiv T {\partial P}/{\partial T}$ are respectively the pressure and enthalpy of the fluid,  
and $u_{\mu}$ is the four velocity field of the plasma, related to the velocity $\textbf{v}$ through 
\begin{equation}
    u_\mu = \frac{\left ( 1, \textbf{v} \right )}{\sqrt{1 - \textbf{v}^2}} = \left ( \gamma, \gamma \textbf{v} \right ) .
\end{equation}
Conservation of the total energy-momentum tensor reads
\begin{equation} \label{eq:: tot_en-mom_cons}
    \partial^\mu T_{\mu \nu} = \partial^\mu T_{\mu \nu}^{\text{plasma}} + \partial^{\mu} T_{\mu \nu}^\phi = 0.
\end{equation}
If we consider a system where the wall has reached a constant velocity $v_w$ along the $z$-direction and we work in its rest frame, then \cref{eq:: tot_en-mom_cons} reads
\begin{equation}
    \label{eq.en_mom_constant}
    \partial_z T^{zz} = \partial_z T^{z0} = 0,
\end{equation}
which, after being integrated across the phase transition front, becomes 
\begin{equation} \label{eq. match_cond}    \gamma^2_+ v_+^2 \omega_+ - \mathcal{F}_+ = \gamma^2_- v_-^2 \omega_- - \mathcal{F}_-, \hspace{2cm} \gamma_+^2 v_+ \omega_+ = \gamma_-^2 v_- \omega_-,
\end{equation}
where $+$ (resp. $-$) denotes quantities in the symmetric (resp. broken) phase. 
Notice that the kinetic term for the scalar field $\phi$ is not present in these conditions, since $\partial_z \phi = 0$ away from the wall, and that the zero temperature potential term $V_0(\phi)$ has been included in the free energy of the system $\mathcal{F} \equiv V_0(\phi) - P$.

These are the so-called matching conditions for the various quantities across the wall. Looking at the case of a plasma described by a bag equation of state, 
\begin{equation} \label{eq. bag_presure}
    \mathcal{F}_\pm = \epsilon_\pm - \frac{a_\pm}{3} T_\pm^4 , 
\end{equation}
and 
\begin{equation} \label{bag_enthalpy}
    \omega_\pm = \frac{4 a_\pm}{3}T_\pm^4,
\end{equation}
we can see from \cref{eq. match_cond}, \eqref{eq. bag_presure} and \eqref{bag_enthalpy} that a change in the bag constant $\epsilon$ between the two phases will result in a change in the plasma temperature and velocity, so that $T_+\neq T_-$ and $v_+ \neq v_-$. For this equation of state, the matching equations can be recast as
\begin{equation}
v_+ = \frac{1}{1+\alpha_+} \left[
\left(\frac{v_-}{2} + \frac{1}{6 v_-} \right) \pm 
\sqrt{
\left(\frac{v_-}{2} + \frac{1}{6 v_-} \right)^2 
+ \alpha_+^2 +\frac23 \alpha_+ - \frac13 }
\right] \, ,
\label{eq:vpvm}
\end{equation}  
where $\alpha_+ = (\epsilon_+ - \epsilon_-)/(a_+ T_+^4)$ parametrizes the strength of 
the phase transition in analogy to the force term in the Boltzmann
equation.  This hydrodynamic relation already displays the origin of the 
singularity~\cite{Dorsch_2022}. When the limit $v_- \to 1/\sqrt{3}$ is taken,
the result is 
\begin{equation}
v_+ \to \frac{1}{1+\alpha_+} \left[
\frac{1}{\sqrt{3}} \pm 
\sqrt{
 \alpha_+^2 +\frac23 \alpha_+ }
\right]
\end{equation}
and hence finite. However, after linearizing (\ref{eq:vpvm}) in $\alpha_+$ this limit becomes singular, just as (\ref{eq. constraint-vwall}). This emphasizes the idea hinted above regarding the breakdown of the standard linearization of the Boltzmann equation: the fluid is collectively changing the value of its temperature and velocity between the two phases. If we expand the particles fluctuations as $T(z) = \Bar{T} + \delta T (z)$ and $v(z) = \Bar{v} + \delta v(z)$ while keeping $\Bar{T}$ and $\Bar{v}$ fixed, then the change in the background temperature and velocity is absorbed in the fluctuations $\delta T (z)$ and $\delta v(z)$ of the light species. In other words, the fluctuations of the light species are not describing non-equilibrium effects only, but some equilibrium effects as well. And these effects will become singular when the Boltzmann equations are linearized.

\section{A new way of solving the Boltzmann Equations} 
\label{sec. new_way_Boltz}

As stated in \cref{sec. lin.BEs}, the main reason why the singularity appears is that the macroscopic change in the hydrodynamic quantities $T_{\text{bg}}$ and $v_{\text{bg}}$ is absorbed in the fluctuations of the light elements. To avoid this, we model the background to be space-dependent, i.e, for a wall moving along the $z$-direction, $T_{\text{bg}} \equiv T_{\text{bg}}(z)$ and $v_{\text{bg}} \equiv v_{\text{bg}}(z)$. The fluctuations of the particles in the plasma will then be written using the fluid Ansatz (\ref{eq:fluid}) but this time with the $z$-dependent 
background.

We define the background by imposing the conservation of its energy-momentum across the phase transition wall. This means solving non-linearly~\cref{eq.en_mom_constant} across the phase transition front, including the contribution of the scalar field $\phi$. This gives the \textit{modified} matching conditions
\begin{equation}\begin{split} 
    v_{\text{bg}}^2 \gamma_{\text{bg}}^2 \omega_{\text{bg}} - \mathcal{F}_{\text{bg}} + \frac{1}{2} (\partial_z \phi)^2 &= \text{k}_1,  \\
    v_{\text{bg}} \gamma_{\text{bg}}^2 \omega_{\text{bg}} &= \text{k}_2.
    \label{eq.1st eq of the system}
\end{split}\end{equation}

The subscript `bg' implies we are considering these equations for the background only, i.e. neglecting out-of-equilibrium contributions. The constants $\text{k}_1$ and $\text{k}_2$ can be computed from the values of temperature and velocity well inside or well outside the bubble, depending on whether the bubble expands as in a deflagration, a detonation, or a hybrid, as we will discuss in the next section.

We can now go back to linearize \cref{eq. BEq} with this new parametrization of the fluctuations. Note that now the Liouville term in the Boltzmann equation will bring in new contributions to the source terms due to the explicit spatial dependence of the temperature and velocity in the equilibrium distributions $f_i^{\text{bg}}$, namely
\begin{equation} \label{eq:: new_sources_def}
    p^\mu \partial_\mu f_i^{\text{bg}} \left( x,p \right)  \supset (f_i^{\text{bg}})' \frac{p^\mu p^\nu}{T} \left( u_\nu \frac{\partial_\mu T}{T} - \partial_\mu u_\nu \right).
\end{equation}

To solve the system of ODEs we only need to repeat the same procedure carried out in \cite{Moore:1995si, Konstandin:2014zta, Dorsch_2022, dorsch_bubble_2023} taking into account these new \textit{hydrodynamic forces}. But for the light species we already saw that there is an equivalent way to solve the system, this time relying on the fact that the two equations for the light degrees of freedom can be fixed thanks to energy momentum conservation. In this work we use~\cref{eq. linearized_en_mom_cons} to find the fluctuations of the light elements, and solve~\cref{eq. Boltz. eq heavy guys OLD} for the top quark and the weak gauge bosons.

 \cref{eq. linearized_en_mom_cons} is a system of two equations. If we focus on its right hand sides, these are respectively given by (see \cref{app: sources_en_mom} for details) 
\begin{equation}
    \label{eq:source1}
    \mathcal{S}_1 = \gamma_w v_w \sum_{i} \partial_z m_i^2 N_i - \gamma_w \frac{4}{3} a T^4 \left( 3 v_w \frac{\partial_z T}{T} + \gamma^2_w \partial_z v \right), 
\end{equation}
where the $\sum_i$ denotes the sum over all the heavy degrees of freedom, and 
\begin{equation}
    \mathcal{S}_2 = - \gamma_w \frac{4}{3} a T^4 \left( \frac{\partial_z T}{T} + \gamma^2_w v_w \partial_z v \right).
\end{equation}

These two equations encode the dynamics of the total energy-momentum tensor and hence 
should be equivalent to (\ref{eq.1st eq of the system}). This means that once we enforce the correct background, these sources 
will exactly vanish and the sources of other moments of the Boltzmann equation will be modified. This should solve the issue with the singularity since 
the problematic equations without damping will not be sourced anymore.
However, there are several caveats. 

First, the Boltzmann equations constitute the energy-momentum of the fluid
and to show the equivalence to (\ref{eq.1st eq of the system}) requires to solve also the equation of motion of the scalar field. 
In our framework, we will assume for the Higgs field a specific shape and then 
iteratively find the correct wall velocity and wall thickness. 
Second, we will neglect the mass dependence in most momenta of the 
Boltzmann equation while we keep the mass dependence in (\ref{eq.1st eq of the system}).

Both of these effects will lead to the fact that the sources in (\ref{eq.1st eq of the system}) are not exactly zero.
However, we will enforce these sources to vanish and then study their
behavior a posteriori, see Sec.~\ref{sec:cancellation}. After that, the two equations 
representing the energy-momentum tensor has then neither a source nor a damping 
and will yield an algebraic relation between the fluctuations in the light and heavy species, see App.~\ref{appendix:: linearized BEs}.

\section{Solutions to the hydrodynamical equation\label{sec. hydrodyn.}}

In order to solve equations~\eqref{eq.1st eq of the system} one must first fix the constants k$_1$ and k$_2$ by establishing appropriate boundary conditions for the temperature and plasma velocity in the vicinity of the bubble wall.  

One defines the nucleation temperature $T_n$ as the temperature at which the bubbles actually start to form. We compute this quantity using standard techniques found in the literature~\cite{Hindmarsh:2020hop, Coleman:1977py, Linde:1981zj}, namely solving the sphaleron equation to find the so-called ``bounce configuration'', calculating its $3$-dimensional Euclidean action $S_E(T)$ and imposing
\begin{equation}
    \dfrac{S_E(T_n)}{T_n} \approx 140,
    \label{eq:S_T}
\end{equation}
which is roughly equivalent to the condition of one bubble per Hubble horizon if the transition takes place at the electroweak scale. Clearly, far away from the bubble the plasma must be at this temperature, since no information on the bubble propagation has yet arrived. But what actually happens at the vicinity of the bubble wall depends on the wall velocity, and there are actually three different classes of solutions: deflagrations, hybrids and detonations~\cite{Espinosa:2010hh}. 

\subsubsection*{Detonations}

These solutions correspond to the case when the plasma immediately ahead of the bubble is still at rest, meaning it has no prior information about the bubble’s arrival, while a rarefaction wave develops behind the wall and drags the fluid. As a result, the temperature just in front of the bubble, $T_+$, equals the temperature far away, $T_n$, and the same holds for the fluid velocity, $v_+ = v_w$.

This situation takes place when the wall velocity exceeds the so-called Jouguet velocity $v_J$, which is in turn determined from the Chapman-Jouguet condition~\cite{PhysRevD.25.2074,Giese_2020} that immediately behind the wall the fluid moves (relative to the wall frame) at the speed of sound, i.e. $v_- = c_{s,b}$. The speed of sound in the plasma is computed from 
\begin{equation}
    c_s^2 \equiv \frac{dP/dT}{de/dT},
\end{equation}
where $e \equiv \omega - P$, and the subscript $b$ indicates that we have to consider its value in the broken phase, which in general differs from the one in the symmetric phase, given by $c_{s,s} = 1/\sqrt{3}$. 
Mathematically, the condition for detonations is expressed as~\cite{Giese_2020}
\begin{equation}
v_w > v_J = \frac{1 + \sqrt{3\alpha_n \left ( 1 - c_{s,b}^2 + 3 c_{s,b}^2 \alpha_n \right )}}{1/c_{s,b} + 3 c_{s,b} \alpha_n},
\end{equation}
where we defined the optimal strength parameter $\alpha_n$ to be \cite{PhysRevD.25.2074,Giese_2020} 
\begin{equation} \label{eq:: definition of alpha_n}
        \alpha_n = \frac{D \bar \theta (T_n)}{3 \omega_n},
\end{equation}
with $\bar \theta = e - {P}/{c_{s,b}^2}$ and $D\bar\theta\equiv \bar\theta_s(T_n) - \bar\theta_b(T_n)$.

\subsubsection*{Deflagrations}
Deflagrations, in contrast, occur when the wall propagates at subsonic speeds, $ v_w < c_{s,b}$. Here, the bubble is preceded by a shock wave that heats up the plasma. The velocity of the shock front, $\xi_\text{sh}$, can be determined using the continuity equations \eqref{eq. match_cond} \cite{Leitao_2011}. The plasma behind the shock moves with velocity $v(\xi_\text{sh})$, and transforming to the shock frame, one can express the plasma velocities as
\begin{equation}
\xi_\text{sh} \, \mu(\xi_\text{sh}, v(\xi_\text{sh})) = \frac{1}{3} \quad \text{and} \quad \frac{\xi_\text{sh}}{\mu(\xi_\text{sh}, v(\xi_\text{sh}))} = \frac{3 T_\text{sh}^4 + T_n^4}{3T_n^4 + T_\text{sh}^4}.
\end{equation}
By solving these, the shock front position is found, and the temperature jump across the shock is determined. For these solutions, in general $v_+ \neq v_w$ and $T_+ \neq T_n$, so one needs to iteratively compute $T_n(T_+, v_w)$ until agreement with the actual nucleation temperature is reached~\cite{Lewicki:2021pgr}. Then, using the fact that for deflagrations $v_- = v_w$ in the wall frame, because the fluid inside of the bubble is at rest with respect to the bubble wall, one can compute $T_-$ and $v_+$ using~\cref{eq. match_cond}.

\subsubsection*{Hybrids}
The third class of solutions is hybrids, which feature supersonic walls ($v_w > c_s$), but differ from detonations by having a shock front ahead of the wall together with the rarefaction wave behind it. Technically, these solutions are also deflagrations, but to be clear we prefer to call them hybrids and reserve ``deflagrations'' for the subsonic case. The method for calculating $T_+$ in hybrids is similar to the one for deflagrations, except that the plasma moves at the speed of sound behind the wall, requiring $v_- = c_s$ in the continuity equations. 

\begin{figure}[t!]
    \centering 
    \includegraphics[width=0.55\textwidth]{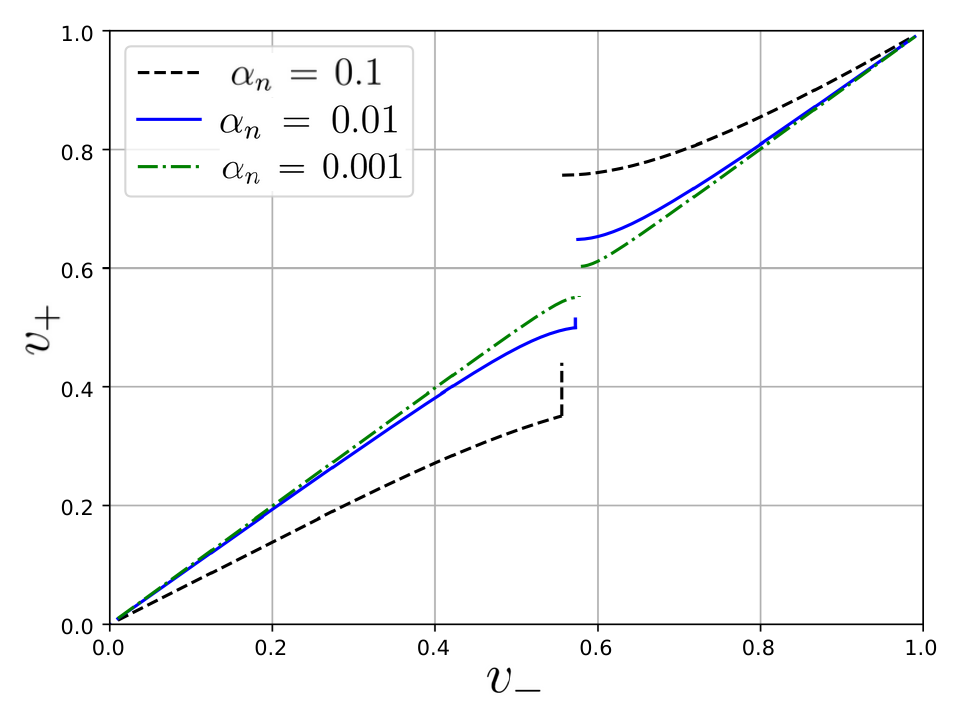}
    \caption{Values of $v_+$ as a function $v_-$ corresponding to solutions of the matching conditions, for different values of the transition strength parameter $\alpha_n$. This plots represents only the solutions where the right boundary condition has been taken into account. Solutions for $v_w<c_{s,b}$ are deflagrations and have $v_+<v_-$. For $c_{s,b}<v_w<v_J$ the solutions are hybrids, corresponding to the vertical lines in the plot, since they have a fixed $v_- = c_{s,b}$. For $v_w>v_J$ one has detonations with $v_+>v_-$. Notice that the point where one passes from deflagrations to hybrids varies with $\alpha_n$ because we are considering a non-constant speed of sound $c_{s,b}$. }
    \label{fig: vplus/vminus}
\end{figure}

\begin{figure} [t]
    \centering 
    \includegraphics[width=0.6\textwidth]{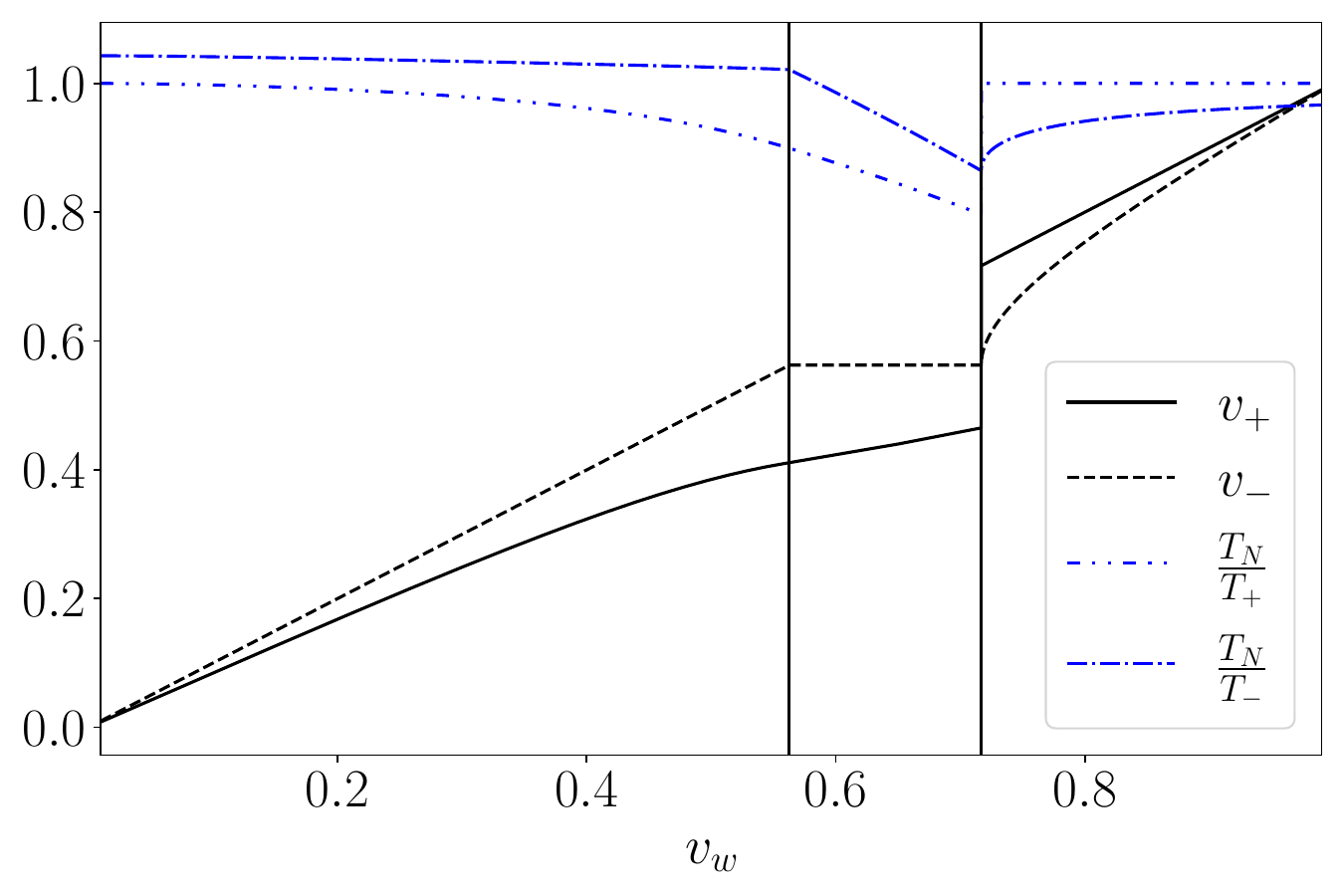}
    \caption{Solutions to the matching conditions for different thermodynamics quantities as a function of the wall velocity $v_w$ for a strength parameter $\alpha_n \simeq 0.05 $. The left vertical line represents the speed of sound $c_{s,b}$, the right one the Jouguet velocity $v_J$. Again the physical boundary conditions have been taken into account, this leading to continuity across $c_{s,b}$ and discontinuity across $v_J$.}
    \label{fig: hydrodynamic quantities}
\end{figure}

Figs.~\ref{fig: vplus/vminus} and \ref{fig: hydrodynamic quantities} show the results of solving the matching conditions for the different hydrodynamical quantities $v_+$, $v_-$, $T_+$ and $T_-$. From both figures we can see that all the quantities are continuous when passing from the deflagration to the hybrid regime, while a discontinuity is present for the quantities in front of the bubble wall, i.e in the symmetric phase, around the Jouguet velocity $v_J$. This is due to the fact that, when crossing this velocity, the fluid in front of the wall suddenly stops being pushed, due to a change in the boundary conditions. So, in hydrodynamics one cannot observe a divergence at the speed of sound $c_s$, while we find discontinuities for some quantities at $v_J$.

\subsection{Temperature and velocity profiles}

Once we know the type of hydrodynamical solution to expect (depending on $v_w$) and the temperature and velocity in front of the bubble are determined, the constants k$_1$ and k$_2$ in~\cref{eq.1st eq of the system}  can be deduced, and one can solve these equations for the profiles of the background temperature and velocity \emph{along} the wall. These will be the $T_{\text{bg}}(z)$ and $v_{\text{bg}}(z)$ that we will plug in the fluid \emph{Ansatz} and in the corresponding system of Boltzmann equations.

\begin{figure}[t!]
    \centering
    \includegraphics[width=0.45\textwidth]{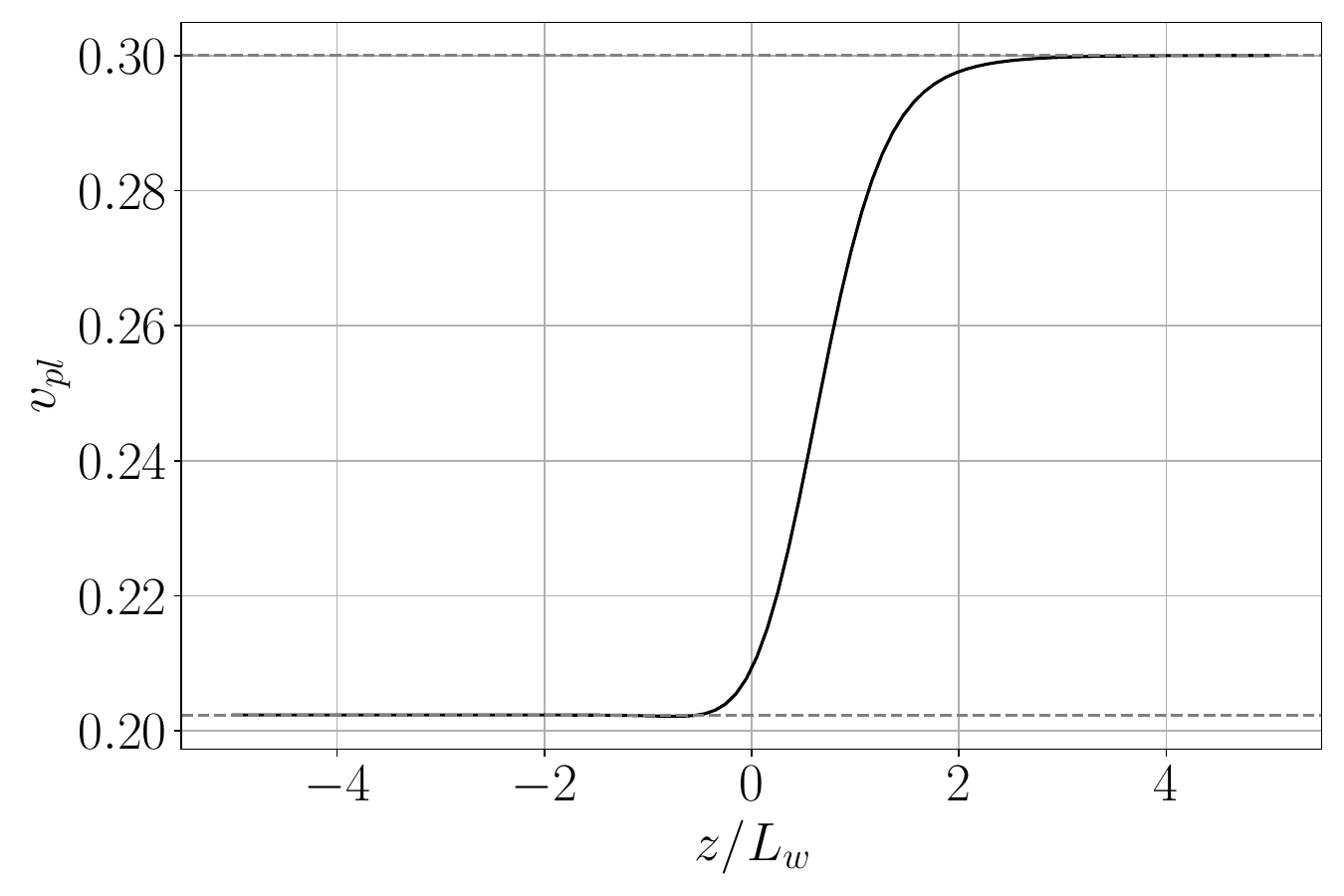}
    \includegraphics[width=0.45\textwidth]{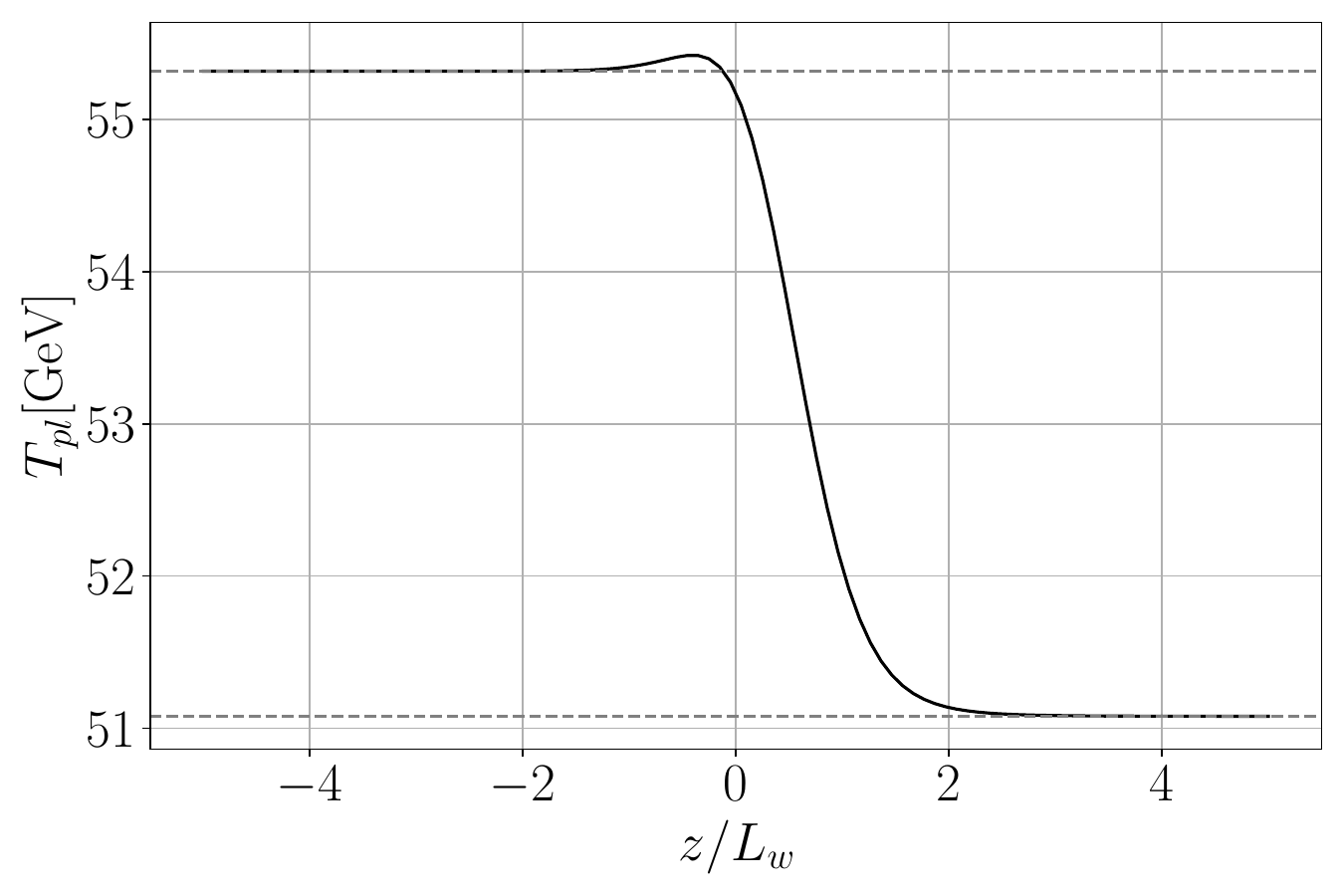}
    \\
    \includegraphics[width=0.45\textwidth]{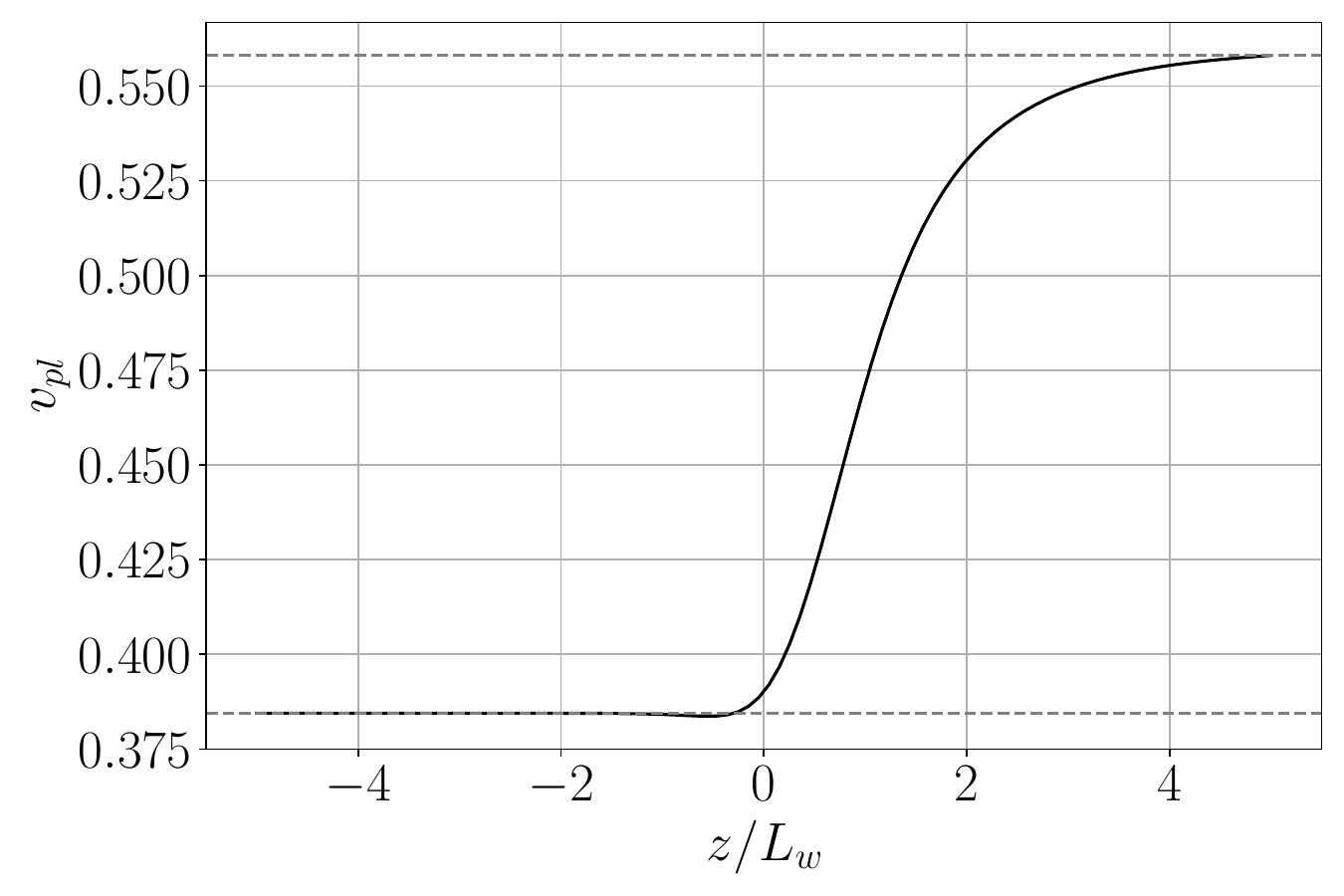}
    \includegraphics[width=0.45\textwidth]{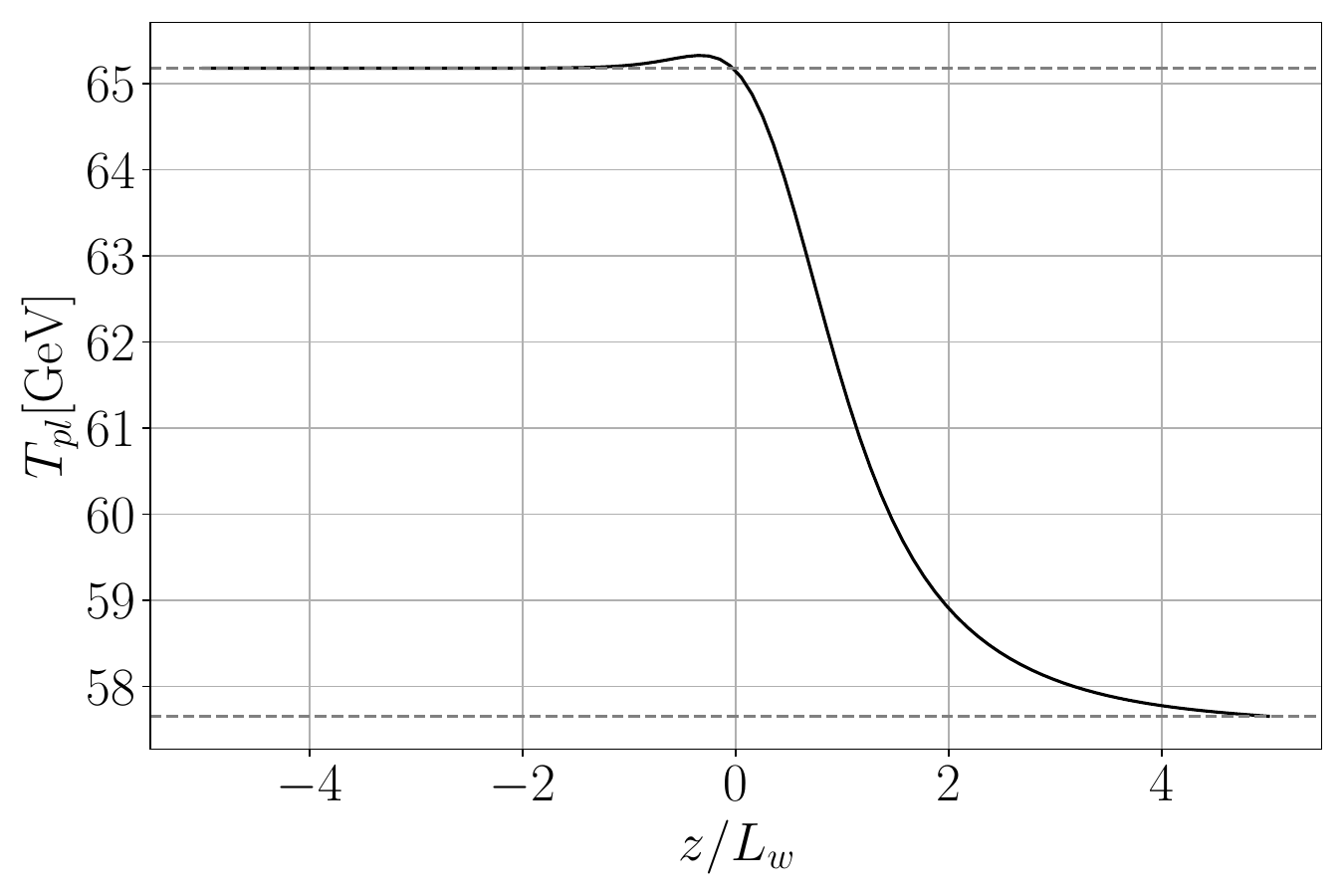}
    \\
    \includegraphics[width=0.45\textwidth]{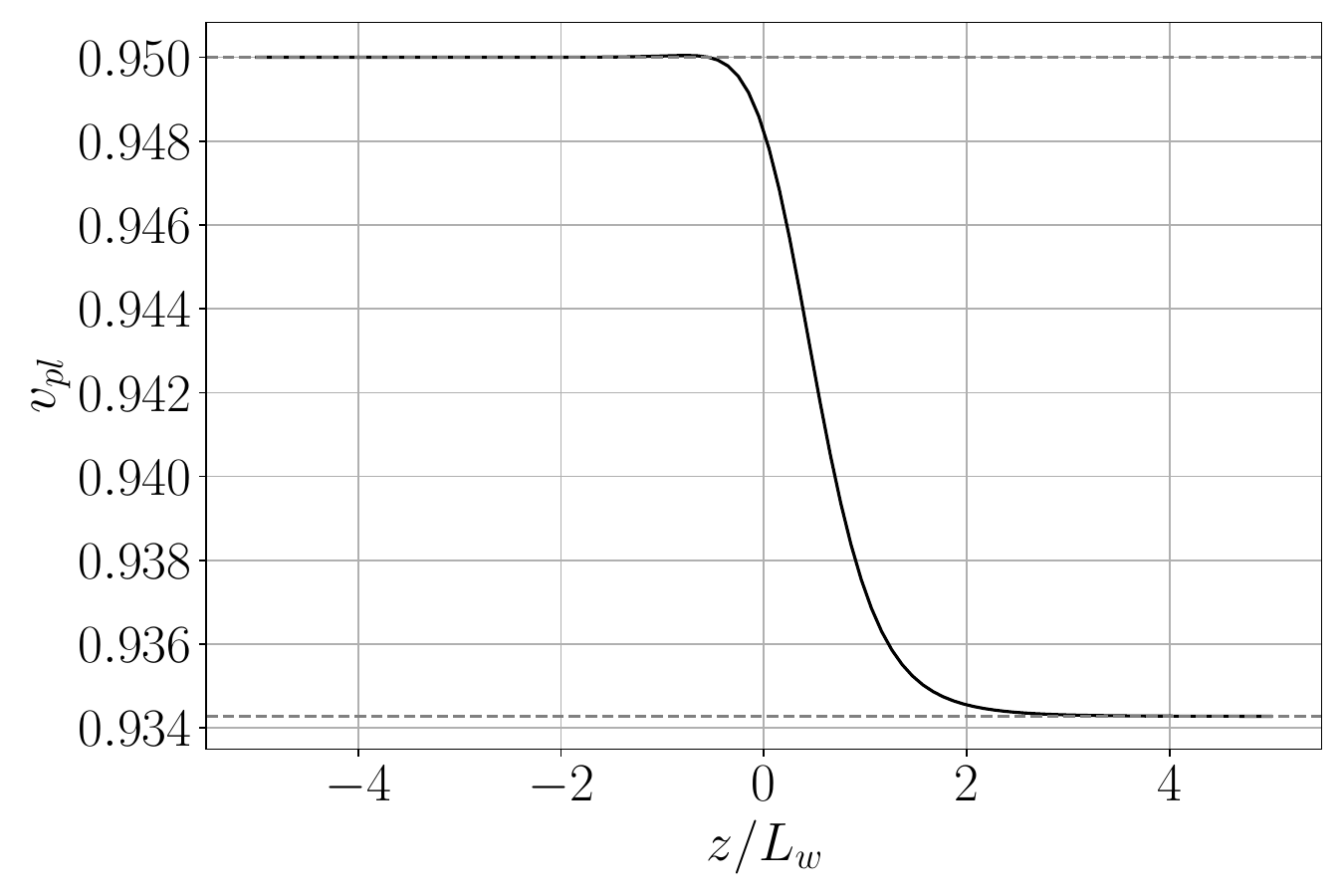}
    \includegraphics[width=0.45\textwidth]{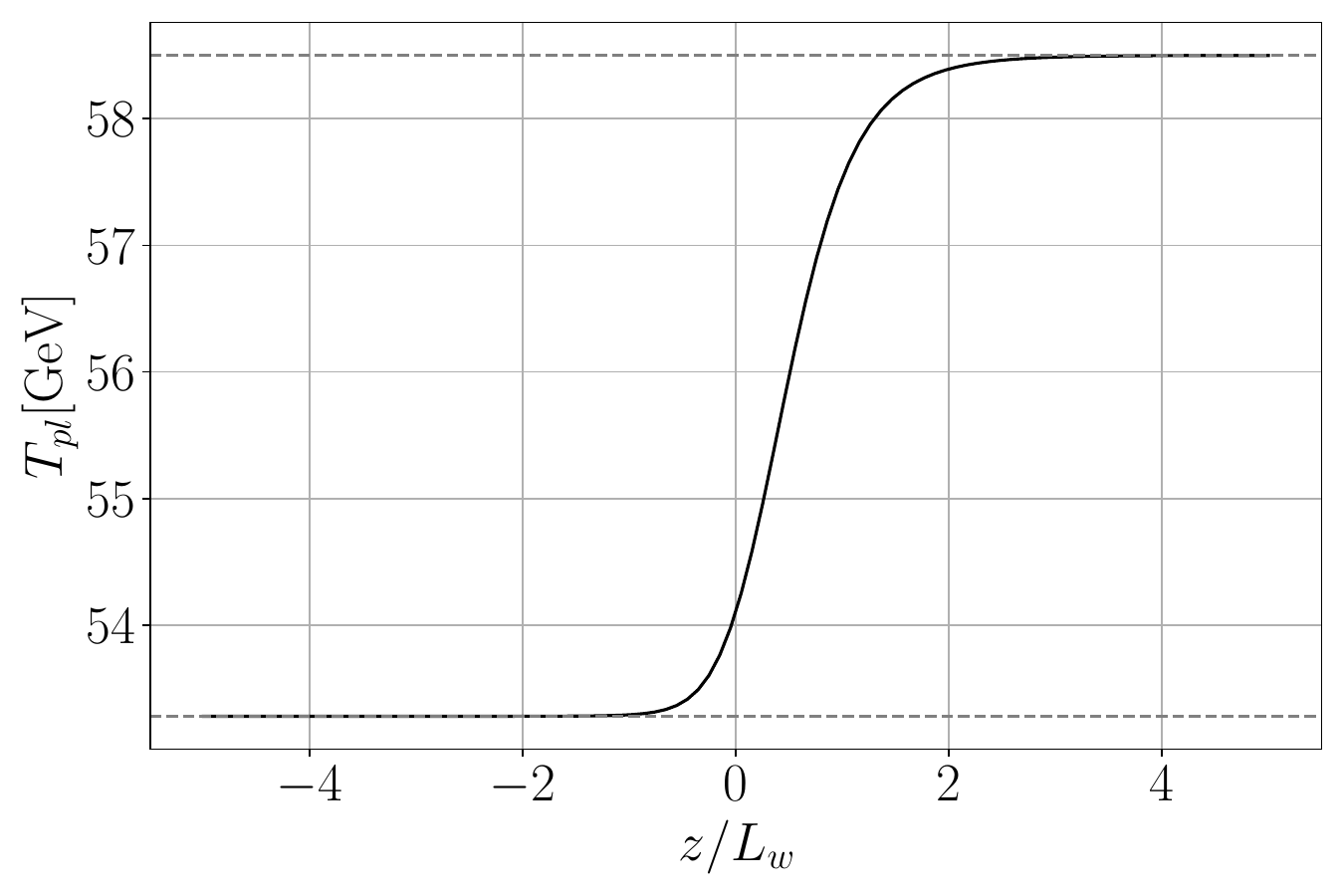}
    \caption{Velocity (left column) and temperature (right column) profiles obtained by solving the non-linear system of equations. The symmetric phase is on the left of the plots. The first row shows the case of a deflagration, the second one a hybrid and the last a detonation.}
    \label{fig:profiles}
\end{figure}

We will solve \cref{eq.1st eq of the system} considering a two-parameter wall-shape \textit{Ansatz}
\begin{equation} 
    \label{eq. wall_shape}
    \phi(z) = \frac{\phi_0}{2} \biggr( 1 + \tanh{\frac{z}{L_w} } \biggr ).
\end{equation}
In \cref{fig:profiles} some temperature and velocity profiles are shown for different types of solutions. These represent the smooth non-linear behavior of the hydrodynamic quantities across the phase transition front. They play an important role in the approach used in this work to solve the linearized Boltzmann system, since they give origin to new source terms, this time not proportional to the change in particles mass, $m m'$. 

\section{The Higgs equation of motion}
\label{sec:: friction}

The passage of the bubble drives the plasma away from equilibrium and this acts on the Higgs field as a counter-pressure against the bubble expansion. Since we want to compute the terminal velocity $v_w$ at which the wall expands, we need to be able to quantitatively account for this pressure. This is usually done by solving the Higgs equation of motion (EOM), which can be deduced from the total energy-momentum tensor conservation in \cref{eq:: tot_en-mom_cons}. In \cite{Konstandin:2014zta} it has been shown that the flux of energy and momentum of the plasma can be computed with the Kadanoff-Baym equations, giving
\begin{equation} \label{eq:: en-mom. flow}
    \partial_\mu T^{\mu \nu}_{pl} = \partial^\nu \phi \sum_i \frac{d m_i^2}{d\phi} \int \frac{d^3 p}{(2\pi)^3} \frac{1}{2E_i} f_i(p^\mu, x^\mu). 
\end{equation}
When putting this together with the energy-momentum of the Higgs field, \cref{eq:: tot_en-mom_cons}, one obtains 
\begin{equation} \label{eq: Higgs_eq_motion}
    \partial_\mu T^{\mu \nu}_{\phi} + \partial^\nu \phi \sum_i \frac{d m_i^2}{d\phi} \int \frac{d^3 p}{(2\pi)^3} \frac{1}{2E_i} f_i(p^\mu, x^\mu) = 0.
\end{equation}
Plugging in the definition $T^{\mu\nu}_\phi$ for a fluid in equilibrium, \cref{eq: Higgs_eq_motion}, one finds
\begin{equation}
    \partial_\mu T^{\mu \nu}_\phi = \partial^\nu \phi \left( \Box \phi + \frac{dV_0}{d \phi} \right), 
\end{equation}
so that the Higgs EOM can be written as
\begin{equation} \label{eq: KG eq_Higgs}
    \Box \phi + \frac{dV_0}{d \phi} + \sum_i \frac{d m_i^2}{d\phi} \int \frac{d^3 p}{(2\pi)^3} \frac{1}{2E_i} f_i(p^\mu, x^\mu)= 0.
\end{equation}
The distribution function can be split into the background (equilibrium) part and the non-equilibrium contribution (from the fluctuations computed via the Boltzmann equation). The equilibrium part can be rewritten in terms of the thermal contribution to the effective potential, thanks to the relation\begin{equation}
    \frac{dP}{dm_i^2} = - \int \frac{d^3p}{(2 \pi)^3} \frac{f_i^\text{bg}(E_i)}{2E_i},
\end{equation}
where the pressure is defined in terms of the equilibrium particle distribution functions through 
\begin{equation}
    P = \frac{1}{3} \sum_i \int \frac{d^3 p}{(2 \pi)^3} \frac{p^2}{E_i} f_i^\text{bg}(E_i).
\end{equation}

Then the Higgs equation of motion \cref{eq: KG eq_Higgs} can be rewritten as~\cite{Dorsch_2022}
\begin{equation} \label{eq: KG Higgs 2}
    - \phi'' + \frac{\partial \mathcal{F}}{\partial \phi} + \sum_i \frac{d m_i^2}{d \phi} \int \frac{d^3 p}{(2 \pi)^3} \frac{1}{2E_i} \delta f_i (\textbf{p}, x) = 0.
\end{equation}
Due to our \emph{Ansatz} for the wall shape in \cref{eq. wall_shape}, there are only two free parameters to be solved for, namely $v_w$ and $L_w$, so we take two moments of \cref{eq: KG Higgs 2}, namely 
\begin{equation} \label{eq: Higgs_momentum_1}
    \int dz [\text{l.h.s of \cref{eq: KG Higgs 2}}] \times \phi' = 0, 
\end{equation}
\begin{equation} \label{eq: Higgs_momentum_2}
    \int dz [\text{l.h.s of \cref{eq: KG Higgs 2}}] \times \phi' ( 2 \phi - \phi_0 ) = 0, 
\end{equation}
These two moments physically represent the vanishing of the total pressure on the wall and of its gradient across it. So, when \cref{eq: Higgs_momentum_1} and \cref{eq: Higgs_momentum_2} are satisfied, the wall is neither accelerating nor stretching. 

\subsection{Fluid backreaction}

\cref{eq: Higgs_momentum_1} can be written explicitly as
\begin{equation} \label{eq: Higgs_ 1st_ integrated}
    \int dz \, \phi' \, \frac{\partial \mathcal{F}}{\partial \phi} + \int dz \, \phi'  \sum_i \frac{d m_i^2}{d \phi} \int \frac{d^3 p }{(2 \pi)^3} \frac{1}{2 E_i} \delta f_i(\textbf{p}, z) = 0.
\end{equation}
The first term corresponds to an equilibrium contribution and can be rewritten as 
\begin{equation} \label{eq: fluid_back}
    \int dz \, \phi' \, \frac{\partial \mathcal{F}}{\partial \phi} 
    = \left. \mathcal{F} \right|^+_- -  \int dz \, T' \, \frac{\partial \mathcal{F}}{\partial T} \,.
\end{equation}
As pointed out in \cite{Konstandin:2010dm, Ai:2021kak, Balaji:2020yrx}, when considering that the temperatures in front and behind the wall are different, there arises a new source of \textit{pressure} on the wall. Since this comes only from equilibrium physics, it is entropy conserving and often referred to as a \textit{backreaction} of the fluid to the wall passing through.
Whether this is enough to stop the wall from accelerating and cause it to reach a stationary configuration depends on the particular model, as will be discussed in \cref{sec. results and discussion}.

\subsection{Out-of-equilibrium contributions} \label{subsection:: out-of-eq friction}
The out-of-equilibrium contributions come from the second term in \cref{eq: Higgs_ 1st_ integrated} and can be rewritten as 
\begin{equation}
\begin{aligned}
    &\frac{N_t}{2 T_+^2} \int dz \,  \frac{d m_t^2}{d z}\left(c_{f 1} \delta \mu_f+c_{f 2} \delta \tau_f+c_{f 2} \delta \tau_\text{light}\right) \,+ \\
    &\frac{N_W}{2 T_+^2} \int dz \,  \frac{d m_W^2}{d z}\left(c_{b 1} \delta \mu_b+c_{b 2} \delta \tau_b+c_{b 2} \delta \tau_\text{light}\right),
\end{aligned}
\end{equation}
where $N_{t,W}$ represents respectively the top and $W^\pm$, $Z^0$ degrees of freedom. We can separate the contributions from the heavy and light species in the plasma by defining
\begin{equation} \label{eq:: f_fl definition}
    f_\text{fl} \equiv \frac{N_t}{2 T_+^2} \int dz \,  \frac{d m_t^2}{d z}\left(c_{f 1} \delta \mu_f+c_{f 2} \delta \tau_f \right) + \frac{N_W}{2 T_+^2} \int dz \,  \frac{d m_W^2}{d z}\left(c_{b 1} \delta \mu_b+c_{b 2} \delta \tau_b\right),
\end{equation}
\begin{equation}
    f_\text{light} \equiv  \frac{N_t}{2 T_+^2} \int dz \, \frac{d m_t^2}{d z} \, c_{f 2} \, \delta \tau_\text{light} + \frac{N_W}{2 T_+^2}  \int dz \, \frac{d m_W^2}{d z} \, c_{b 2} \, \delta \tau_\text{light}.
\end{equation}
The same can be done for the ones in \cref{eq: Higgs_momentum_2}, defining $g_\text{fl}$ and $g_\text{light}$. 
In this way \cref{eq: Higgs_momentum_1} and \cref{eq: Higgs_momentum_2} become
\begin{equation} \label{eq: Higgs_solver1}
    \frac{\mathcal{\left. F \right|_-^+}}{T_+^4} - \frac{1}{T_+^4}\int dz \, T' \, \frac{\partial \mathcal{F}}{\partial T} + f_\text{fl} + f_\text{light} = 0,
\end{equation}
\begin{equation} \label{eq: Higgs_solver2}
    \frac{2}{15(T_+L)^2} \left( \frac{\phi_0}{T_+} \right)^3 + \frac{W}{T_+^5} + g_\text{fl} + g_\text{light}= 0,
\end{equation}
with
\begin{equation}
 W = \int_0^{\phi_0} \frac{\partial{\cal F}(\phi,T)}{\partial\phi} (2 \phi - \phi_0) d\phi \, ,
\end{equation}
and can be solved for $v_w$ and $L_w$.

The whole formalism defined in \cref{sec. new_way_Boltz} and \cref{sec:: friction} offers us the possibility of implementing energy-momentum conservation at non-linear level, see eq.~(\ref{eq.1st eq of the system}). In fact, after having solved for the background velocity and temperature field in \eqref{eq.1st eq of the system} and the Higgs EOM in \eqref{eq: Higgs_solver1}-\eqref{eq: Higgs_solver2}, we naturally obtain that the sources for \cref{eq. linearized_en_mom_cons} vanishes at linear order, as shown  in~\cref{appendix:: source canc.}. 
For this reason, we solve our system enforcing the condition $\mathcal{S}_1 = \mathcal{S}_2 = 0$ and we check and discuss the validity of this approximation in \cref{sec. results and discussion}.
This condition, which is even more severe than \cref{eq. linear-en-mom}, means that the fluctuations  of the light degrees of freedom vanish away from the wall for every value of the wall velocity and that the singularity in the friction is resolved.

\subsection{A benchmark model: the SM with a low cutoff} 
\label{sec. benchmark_mod}

As a benchmark model we choose a simple extension of the SM with an effective $\phi^6$ operator, representing new physics coming into play at a scale $\Lambda$. This framework allows for a first-order phase transitions with a Higgs mass compatible with
present LHC data~\cite{Grojean:2004xa}. The new physics is supposed to contribute to the Higgs potential but does not contribute significantly to the friction, so we keep modeling the plasma as containing only SM particles. 
In this case the high temperature expansion of the Higgs
effective potential can be written as
\begin{equation}
    V(\phi, T) = \mathcal{F}(\phi, T) = V_0(\phi) - \frac{a}{3} T^4 + \frac{c}{2} \phi^2 T^2, 
\end{equation}
with
\begin{equation}
    a = 106.75 \dfrac{\pi^2}{ 30},
    \quad\text{and}\quad
    c = \frac{1}{16} \biggr(4 y_t^2 + 3 g^2 + 4 \frac{m_H^2}{v^2} - 12 \frac{v^2}{\Lambda^2} \biggr),
\end{equation}
and $V_0(\phi)$ the zero temperature potential given by 
\begin{equation} \label{eq. zero-temp-pot}
    V_0(\phi) = - \frac{\mu^2}{2} \phi^2 + \frac{\Tilde{\lambda}}{4} \phi^4 + \frac{1}{8 \Lambda^2} \phi^6.
\end{equation}
The parameters of this potential are set in such a way that we recover the measured SM values for the Higgs mass $m_H = 125$ GeV and the Higgs VEV $v=246.22$ GeV, yielding
\begin{equation}
    \mu^2 = \frac{m_H^2}{2} - \frac{3}{4} \frac{v^4}{\Lambda^2}, 
    \qquad
    \Tilde{\lambda} = \frac{m_H^2}{2 v^2} - \frac{3}{2} \frac{v^2}{\Lambda^2}. 
\end{equation}
\noindent It should be noted here that, in contrast to the SM scenario, the value of the quartic coupling $\Tilde{\lambda}$ is negative for regions of the parameter space where a first order phase transition occurs, and this is possible thanks to the presence of the $\phi^6$ term which stabilizes the potential. Moreover, we include thermal corrections up to order $\mathcal{O}(m^2 T^2)$ since for these classes of models the first-order nature of the phase transition does not rely on the presence of cubic thermal contributions of the bosons $\propto \phi^3 T$. 

\section{Results and discussions} 
\label{sec. results and discussion}

For this benchmark model we solve the Higgs equation of motion for different values of the parameter $\Lambda \in [600-800]$ GeV, corresponding to different phase transition strengths. Solving the Higgs equation itself demands a solution of the Boltzmann equation, which in turn requires solving the hydrodynamical energy-momentum continuity equation for determining the background profile across the bubble wall. Nevertheless, the whole procedure takes about a few minutes on a desktop computer for each point in the parameter space.

\subsubsection*{Friction functions and total pressure on the wall}
\begin{figure}[t]
    \centering
    \includegraphics[width=0.48\textwidth]{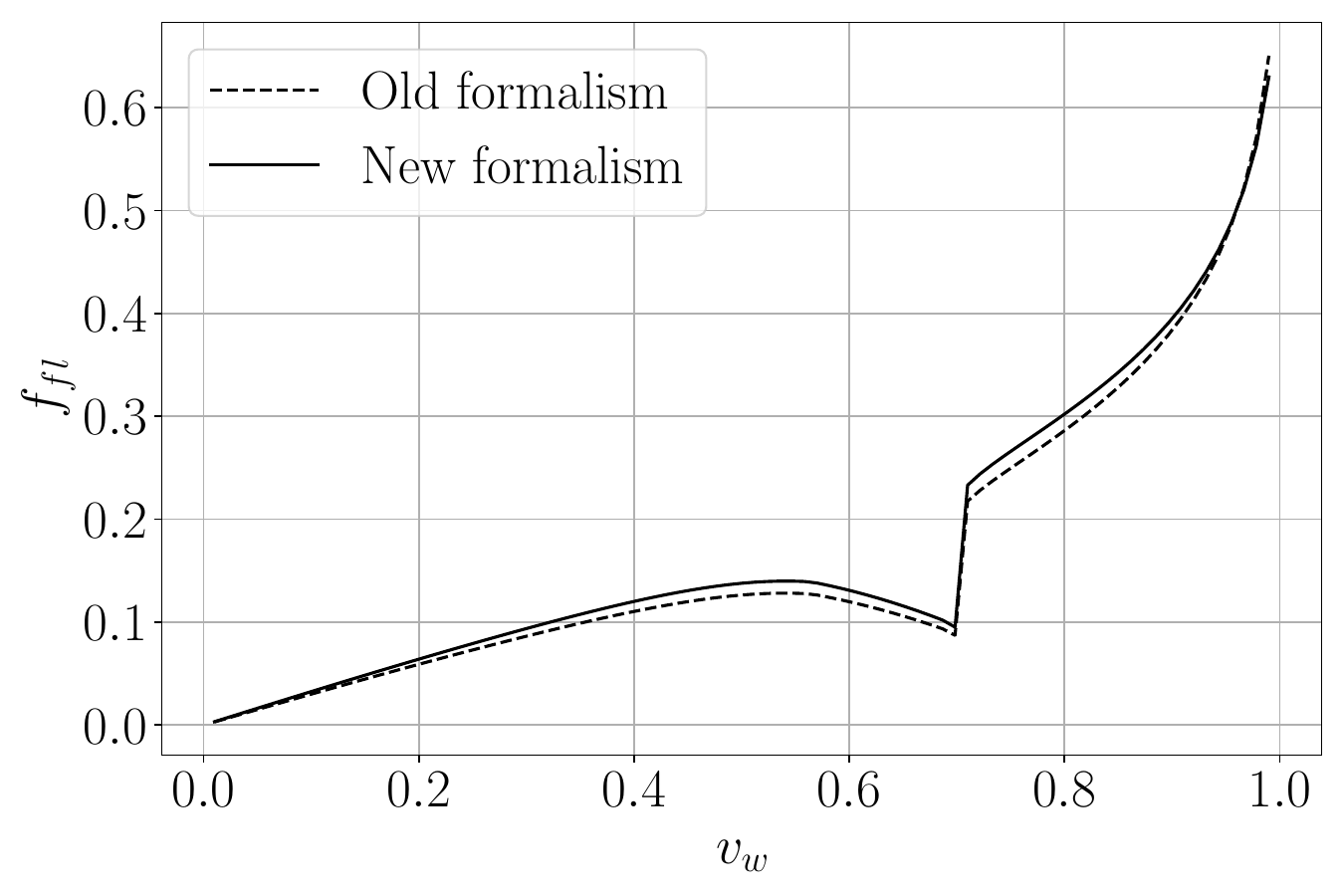}
    \includegraphics[width=0.48\textwidth]{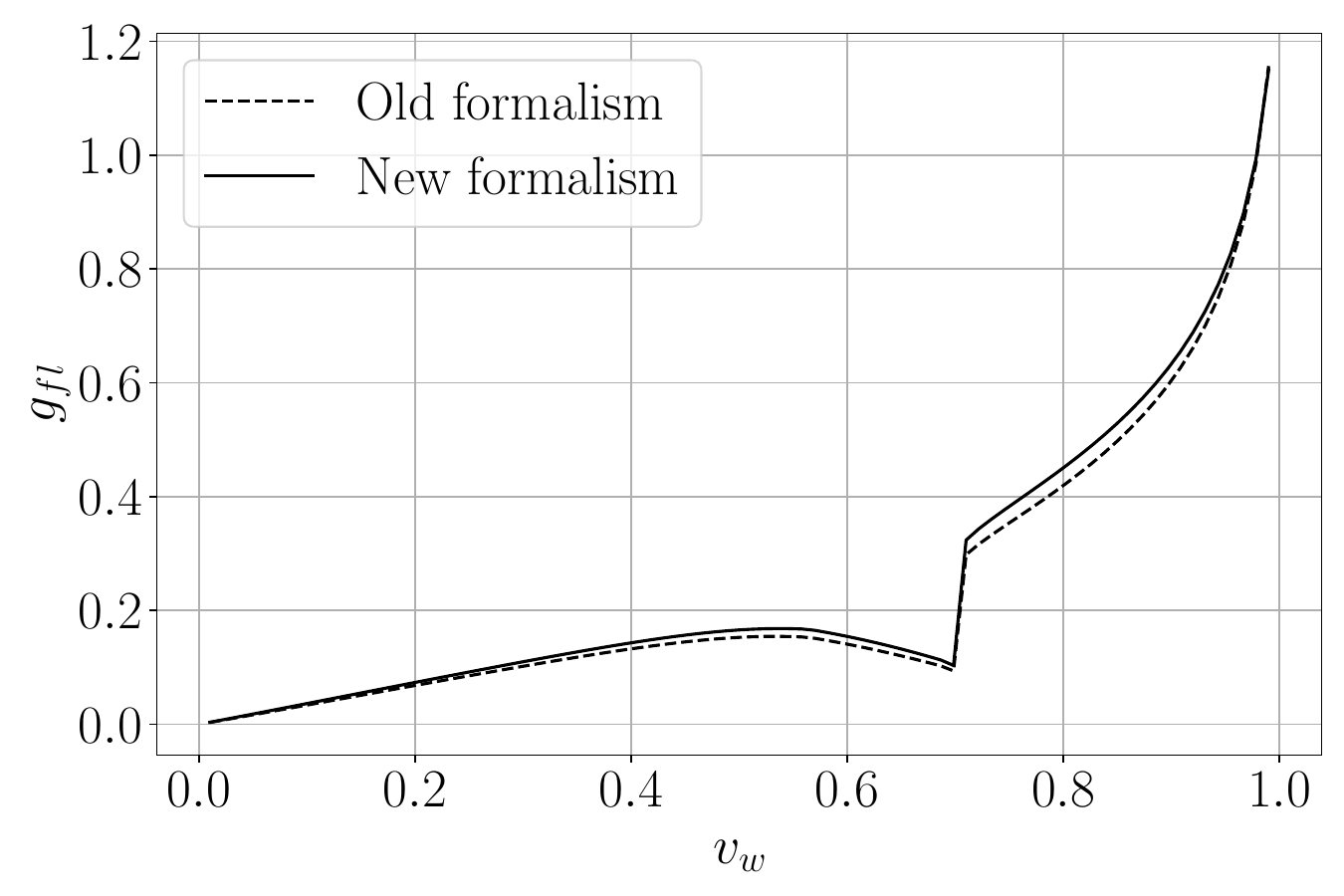}
    \caption{Comparison of the out-of-equilibrium friction from the heavy particles in the plasma ($f_{pl}$ in the left panel and $g_{pl}$ in the right one) found with the new approach with respect to the one found with the old approach. The cutoff scale here is set to $\Lambda = 625$ GeV.}
    \label{fig:frict_comparison}
\end{figure}

\begin{figure} [t]
    \centering 
    \includegraphics[width=.48\textwidth]{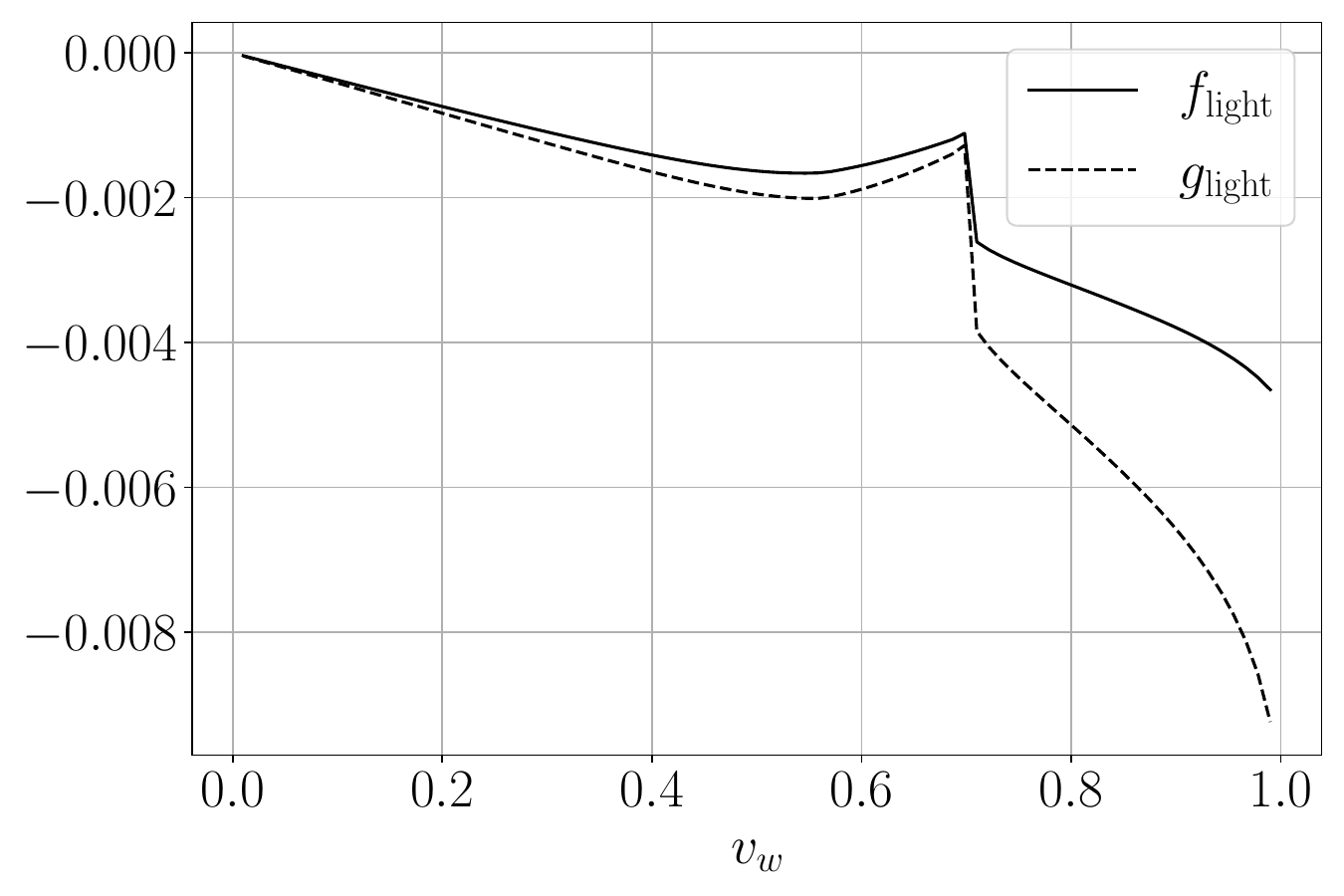}
    \includegraphics[width=.48\textwidth]{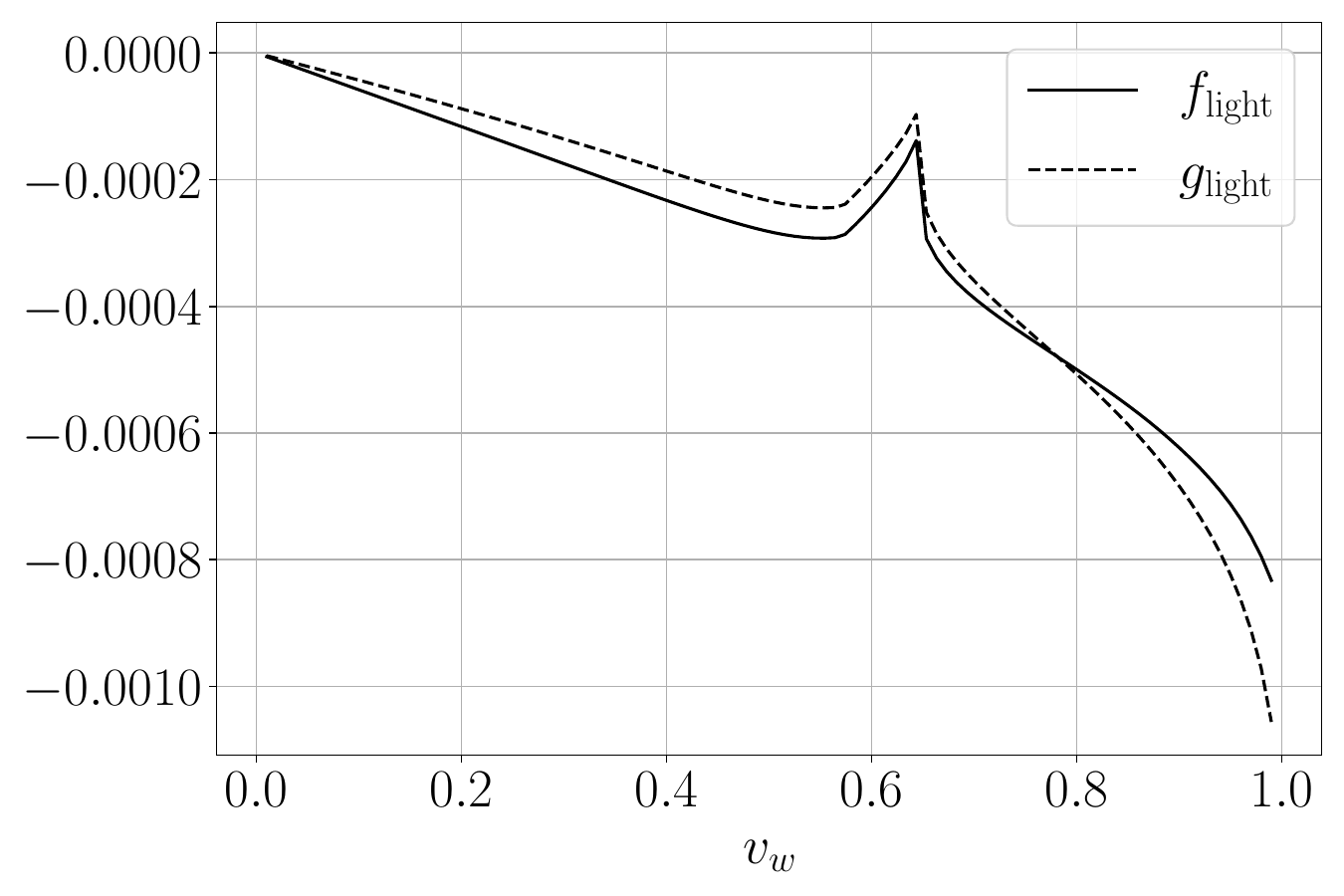}
    \caption{Pressure on the expanding wall coming from the out-of-equilibrium fluctuations of the light particles in the plasma as a function of the wall velocity $v_w$. The values are for a cutoff scale $\Lambda = 625$ GeV (left) and $\Lambda = 690$ GeV (right), which correspond to a value of the strength parameter $\alpha_n = 0.04$ (left) and $\alpha_n = 0.01$ (right). We see that the singularity in the friction is not present any more, while a discontinuity appears at the Jouguet velocity $v_J$, as expected on hydrodynamical grounds.}
    \label{fig: light_friction}
\end{figure}

We first show in \cref{fig:frict_comparison} the out-of-equilibrium contribution to the friction coming from the heavy species in the plasma for $\Lambda = 625$ GeV. The left plot represents the function $f_\text{fl}$ defined in \cref{eq:: f_fl definition}, while the right one is $g_\text{fl}$. This friction is a growing function of the wall velocity for deflagration and detonations regimes, while it decreases with $v_w$ for hybrids. A discontinuity is present at the Jouguet velocity $v_J$ and this is due to the change in the boundary conditions, as discussed in \cref{sec. hydrodyn.}.
The black dashed lines in \cref{fig:frict_comparison} represent the same functions $f_\text{fl}$ and $g_\text{fl}$ but computed neglecting the \textit{new} source terms from \cref{eq:: new_sources_def}, i.e neglecting the $z$-dependence in the background field. We can see that the two functions are very close to each other, meaning that for the heavy particles considered, the main contribution to the friction is due to their gaining mass from the Higgs. 

Having discussed the friction terms due to the heavy species, we now turn to the contribution of the light ones, which are shown in \cref{fig: light_friction} for two different values of the strength parameter $\alpha_n$. The light degrees of freedom were precisely the ones that led to a singular behaviour in friction in refs.~\cite{Konstandin:2014zta, Dorsch_2022, dorsch_bubble_2023} which now disappeared. The smooth behaviour in this approach is due to the fact that, here, the light species are really encapsulating only their \emph{non-equilibrium} fluctuations around the equilibrium background, whereas in refs.~\cite{Konstandin:2014zta, Dorsch_2022, dorsch_bubble_2023} they described both the equilibrium contribution as well as the non-equilibrium fluctuations. 
For the same reason, the light and heavy contributions are proportional to each other.
The difference in sign is due to the fact that these have been obtained trough energy-momentum conservation and, as expected, their magnitude is also suppressed by the number of the light degrees of freedom, see App.~\ref{appendix:: linearized BEs}.

\begin{figure}[t]
    \centering
    \includegraphics[width=.49\textwidth]{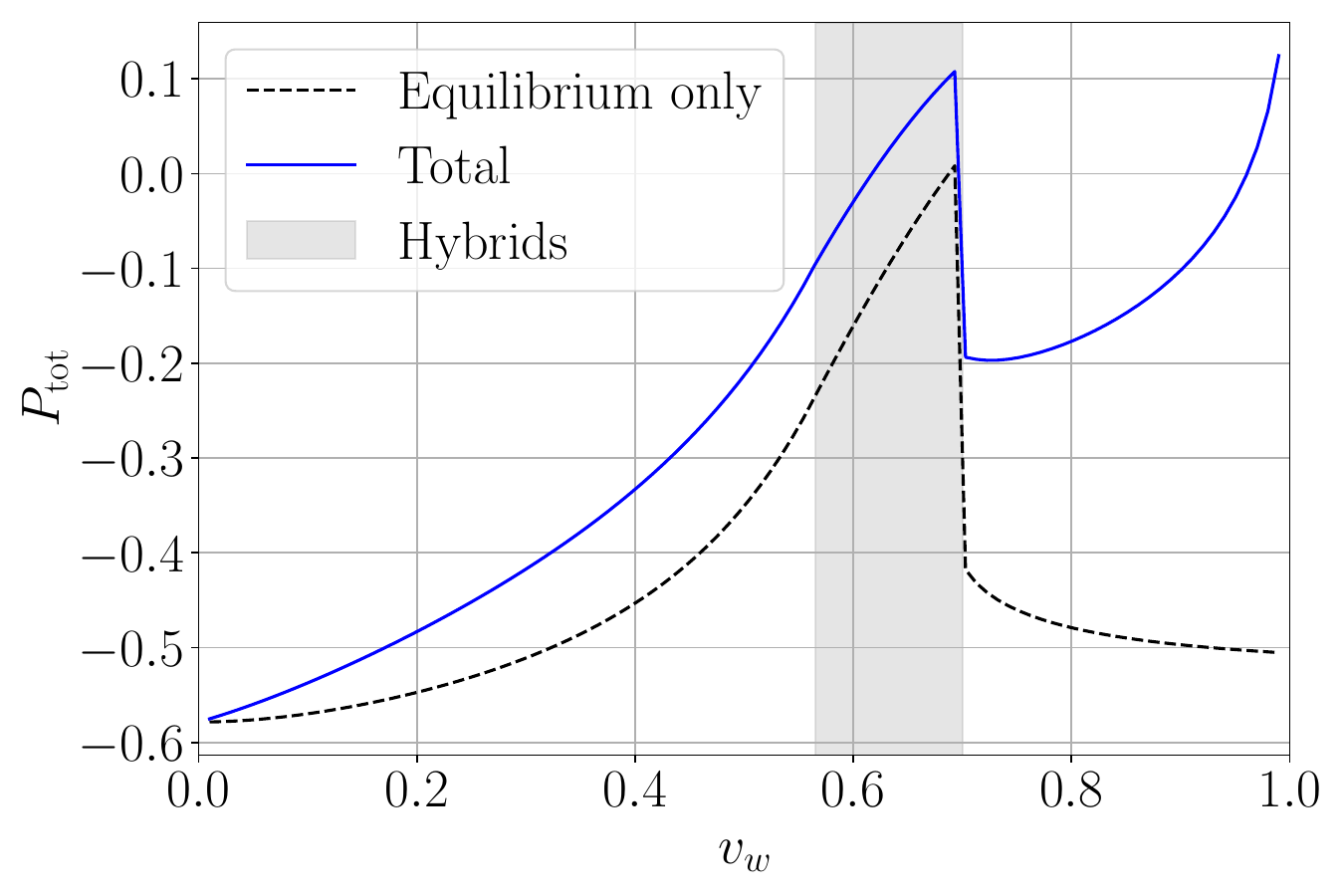}
    \includegraphics[width=.49\textwidth]{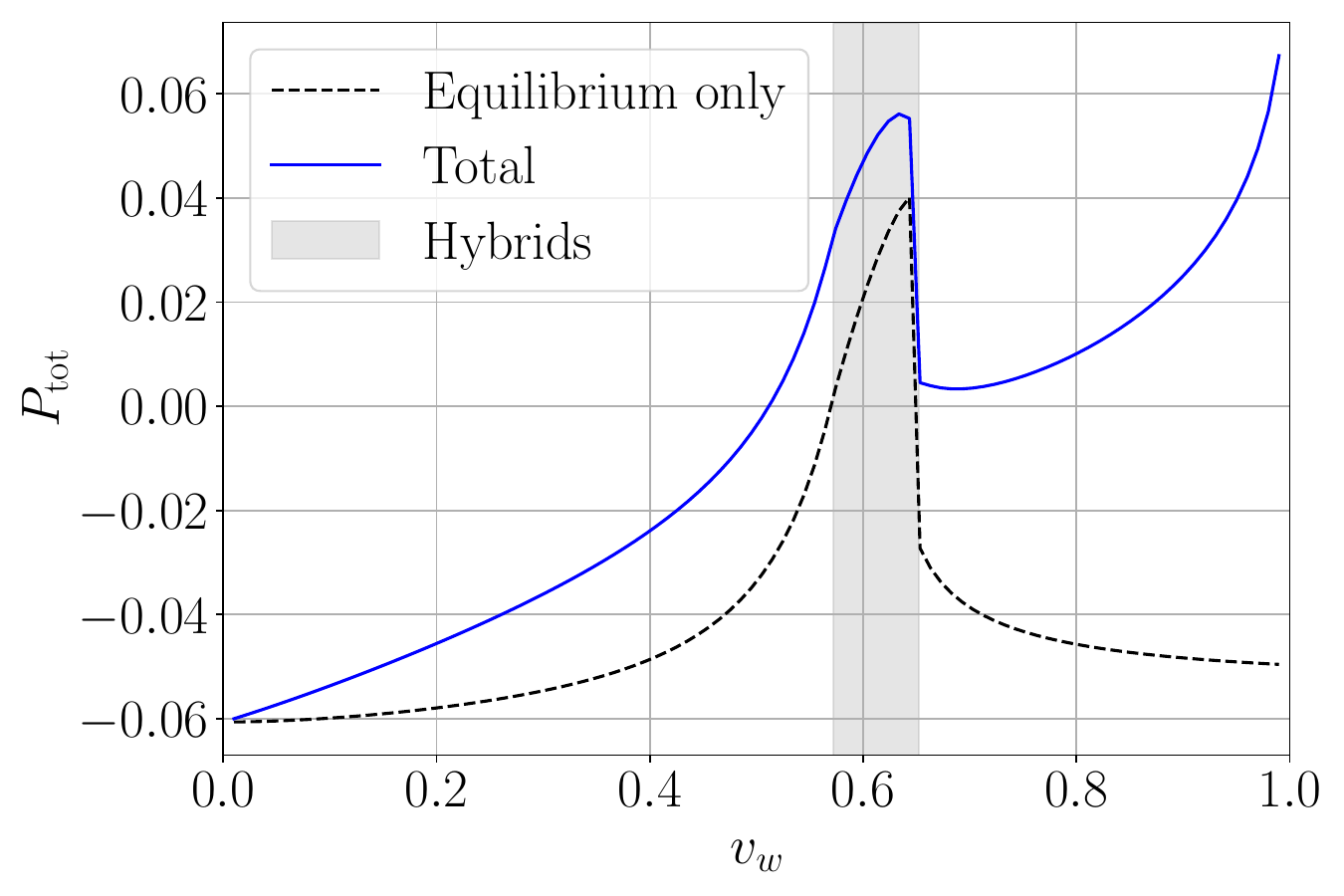}
    \caption{Equilibrium only and total pressure acting on the wall. On the left is represented the case for a cutoff $\Lambda = 625$ GeV, corresponding to $\alpha_n = 0.04$, for which we find both a deflagration and a detonation solution. On the right we show the case $\Lambda = 690$ GeV, with $\alpha_n = 0.01$, for which only a deflagration solution is found. The wall width is fixed in both cases to $L_w T_+ \simeq 15$. 
    }
    \label{fig:pressure on th wall}
\end{figure}

It is also interesting to evaluate the total pressure on the wall, as shown in \cref{fig:pressure on th wall} for the same values of $\alpha_n$ as in~\cref{fig: light_friction}. For comparison, we also show the purely equilibrium contribution to the pressure. The shaded grey region represents the hybrids regime. The stronger the phase transition is, the higher the Jouguet velocity $v_J$, and the larger the hybrid region becomes. We can see here the role of the out-of-equilibrium fluctuations: they increase the pressure acting against the expansion of the wall. 

It has been argued in the literature~\cite{Ai:2021kak,laurent_first_2022,Ai:2023see} that equilibrium considerations would suffice for an adequate evaluation of the pressure against the wall, since the non-equilibrium contribution would be comparably negligible. Our results in~\cref{fig:pressure on th wall} do not seem to corroborate this claim. 

\subsubsection*{Terminal wall velocity and width}

\begin{figure}[t!]
    \centering
    \includegraphics[width=.48\textwidth]{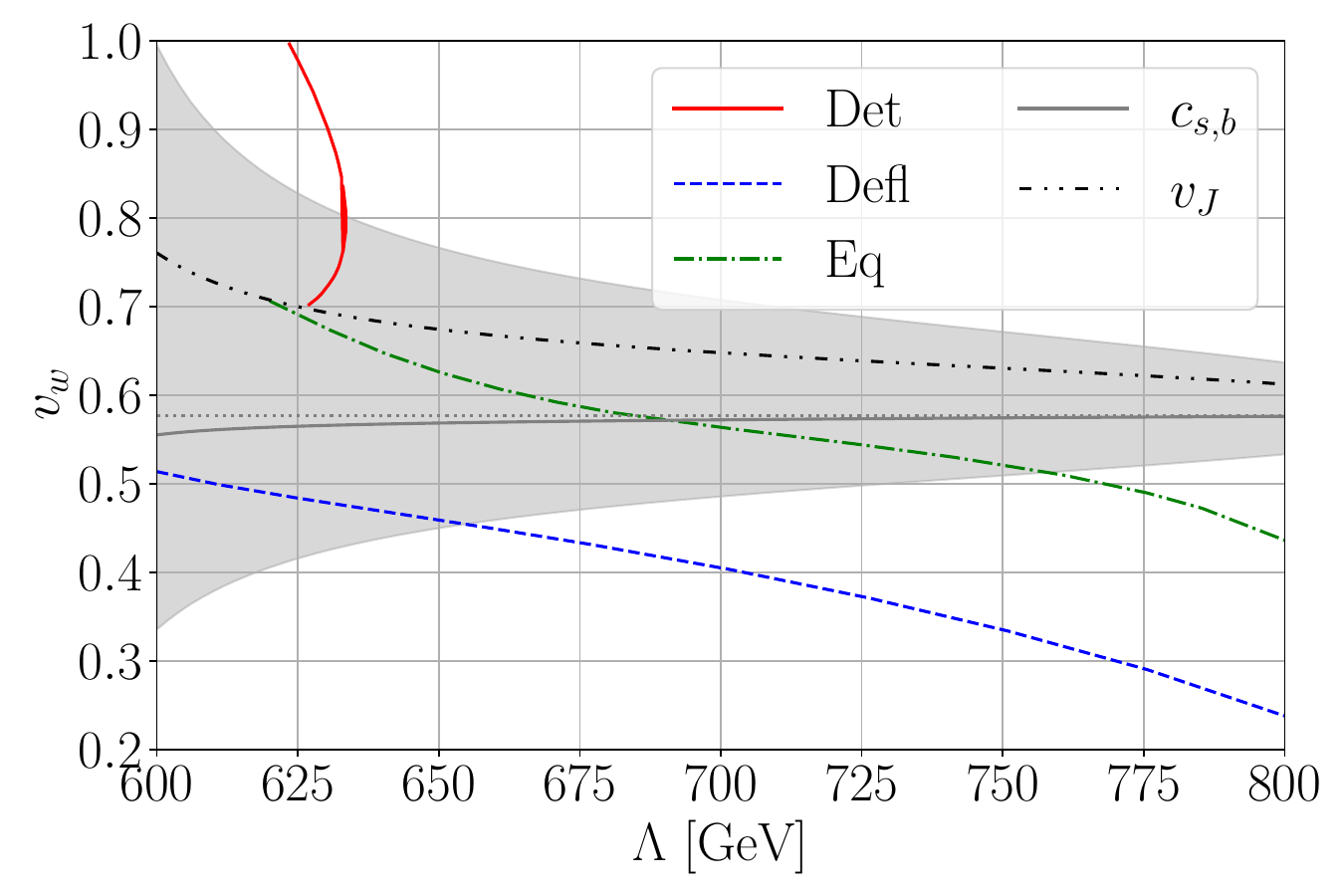}%
    \includegraphics[width=.48\textwidth]{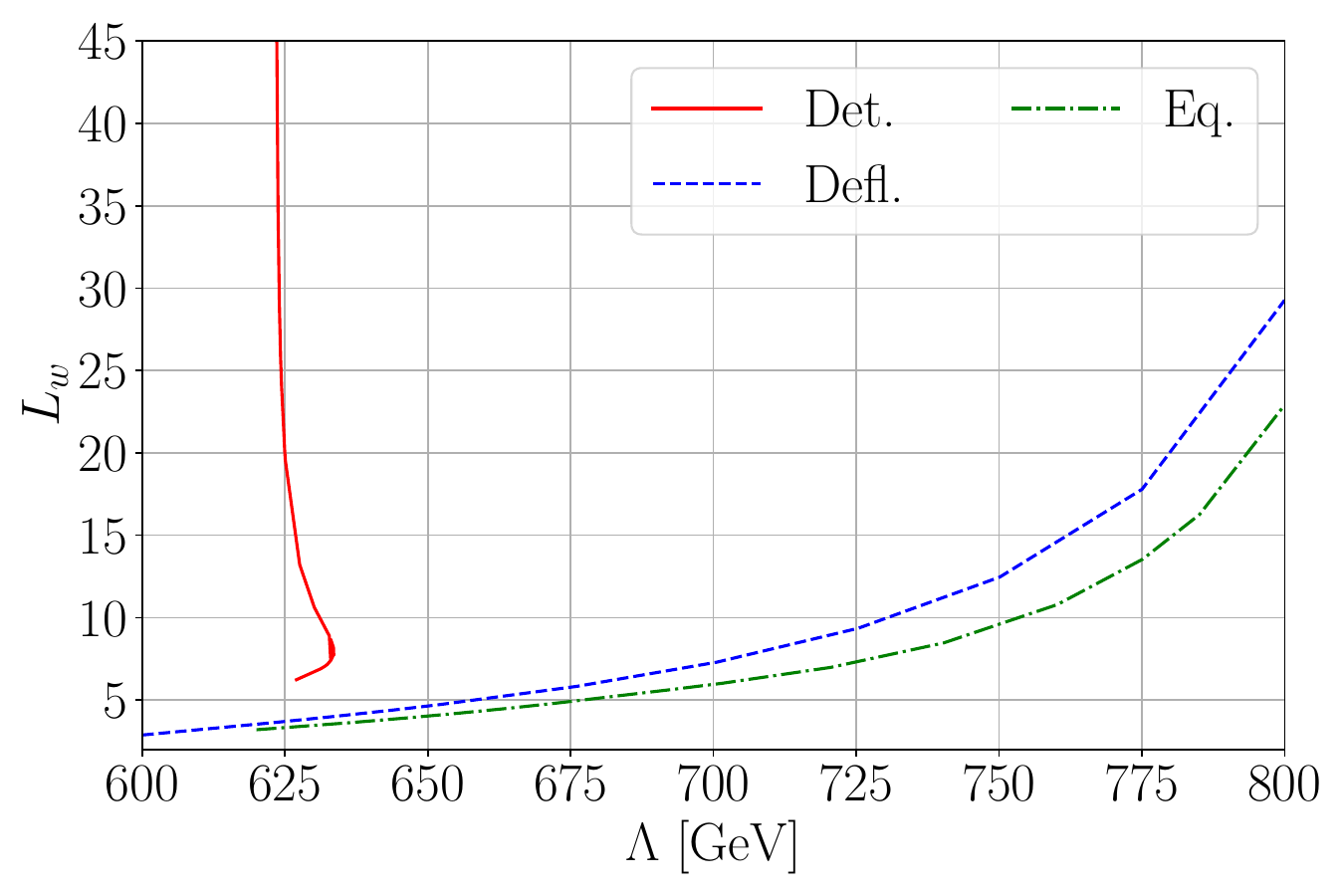}
    \caption{(Left panel) Wall velocities $v_w$, together with the speed of sound $c_{s,b}$ (solid grey line) and the Jouguet velocity $v_J$ (dash-dotted line), in terms of the cutoff scale $\Lambda$. The red line represents detonation solutions, the blue line deflagrations, while the green one shows the values of the wall velocities found neglecting out-of-equilibrium contributions. The grey dotted line represents the speed of sound in the symmetric phase, $c_{s,s} = 1/\sqrt{3}$. The shaded grey region represents the regime where the usual linearization procedure breaks down.  (Right panel) Same but for the wall width $L_w$. Again, red and blue lines take into account all the contributions while the green line is obtained considering equilibrium only.
    }
    \label{fig: solutions to the system}
\end{figure}

To further explore this issue, we analyze the impact of including the out-of-equilibrium terms in the computation of the terminal wall velocity and wall width. The results are shown in~\cref{fig: solutions to the system} as a function of the cutoff scale $\Lambda$. The red lines represent detonation solutions while the blue ones show deflagrations. Green lines represent the solutions found when neglecting the out-of-equilibrium contributions of \cref{subsection:: out-of-eq friction}. In the left panel we also show the speed of sound $c_{s,b}$, which sets the passage from the deflagrations to the hybrids, and the Jouguet velocity, where detonations start. The right panel shows the solutions for the wall width $L_w$ with the same color grading (i.e red for detonations, blue for deflagrations and green for solutions coming from the equilibrium backreaction only).

Notice that, for our specific benchmark model, non-runaway detonations can only be realized in a small range of the parameter space, corresponding to a cutoff scale $\Lambda \in [625 - 635]$~GeV and to strength parameters $\alpha_n \sim 0.04$. This is because, for detonations in the $\Lambda < 625$~GeV region, the counter-pressure is never enough to stop the wall from accelerating. 

On the other hand, deflagrations are always found for all values of $\Lambda$ in the range we explored, and are the only viable solutions for $\Lambda > 635$ GeV. These solutions show a monotonically decreasing behavior, consistent with the fact that higher cutoff scales correspond to weaker phase transitions, i.e. smaller $\alpha_n$. These results are compatible with what has been found in the literature before, see e.g~\cite{Konstandin:2014zta, dorsch_bubble_2023, DeCurtis:2024hvh}, while they slightly differ from the ones in~\cite{Cline:2020jre}, which only include the top quarks in the friction. 

In the range of parameters $\Lambda$ where more than one solution is allowed, one needs to understand which solution will actually be realized. When a bubble is nucleated, it starts accelerating from $v_w = 0$ until reaching the terminal velocity. If the solutions we find are always stable, this means that the only physical solutions will be deflagrations.
For the benchmark model presented here in \cref{sec. benchmark_mod}, the stability of the deflagration solutions is ensured by the behavior of the two momenta in Eqs.~\eqref{eq: Higgs_momentum_1} and \eqref{eq: Higgs_momentum_2}, which are monotonically increasing before the discontinuous jump at $v_J$. 
All detonation solutions we found coexist with a deflagration
solution, as in~\cite{Cline:2021iff, Lewicki:2021pgr}.

\subsubsection*{The importance of adding out-of-equilibrium contributions}

We also show in \cref{fig: solutions to the system} (green lines) the terminal wall velocity and wall width if we took into account only the equilibrium contributions.
We notice that, in this case, the deflagration solutions always correspond to larger wall velocities than the ones found with the full contributions. Looking at \cref{fig:pressure on th wall} one notices that the non-equilibrium terms always increase the pressure against the wall, so neglecting these terms decreases the friction and therefore tends to increase the wall velocity and decrease the wall width. Quantitatively, including out-of-equilibrium contributions gives a correction ranging from $\sim 25 \%$ to $\sim 45 \%$ for $v_w$ and from $\sim 10 \%$ to $\sim 25 \%$ for $L_w$. 

Moreover, we see that no more deflagration solutions can be found for $\Lambda \lesssim 620$ GeV, where $v_w = v_J$, and in this case all detonations are runaways. This can again be understood from \cref{fig:pressure on th wall}. As we increase the phase transition strength, the pressure driving the bubble forwards increases as well. At first this can be equilibrated by the also rising trend in the counter pressure, as seen in \cref{fig:pressure on th wall}, but at $v_J$ the pressure suffers a sudden decrease and becomes monotonically decreasing, if one considers equilibrium-only. In this case, once the Jouguet velocity is overcome, the fluid can no longer exert enough counter-pressure to stop the wall from accelerating. This is precisely the behavior discussed in~\cite{Cline:2021iff, Lewicki:2021pgr}, and which has been assigned to the well-known hydrodynamical obstruction which follows from equilibrium considerations~\cite{Konstandin:2010dm}. However, once we include non-equilibrium terms, the counter-pressure can again increase in the detonation regime, and even overcome the peak at $v_w=v_J$, meaning it could even possibly overcome the inner pressure driving the bubble forwards, thus stopping it from running away. Still, it is worth noticing that this seems to involve some degree of tuning. In the benchmark model we consider, a non-runaway detonation happens only for $\Lambda\in [625-635]$~GeV, as mentioned above. We also stress that the existence of non-runaway detonations is in agreement with recent results in ref.~\cite{Ekstedt:2024fyq}.

\subsection{On-shell source cancellation\label{sec:cancellation}}

\begin{figure} [t!]
    \centering 
    \includegraphics[width=.48\textwidth]{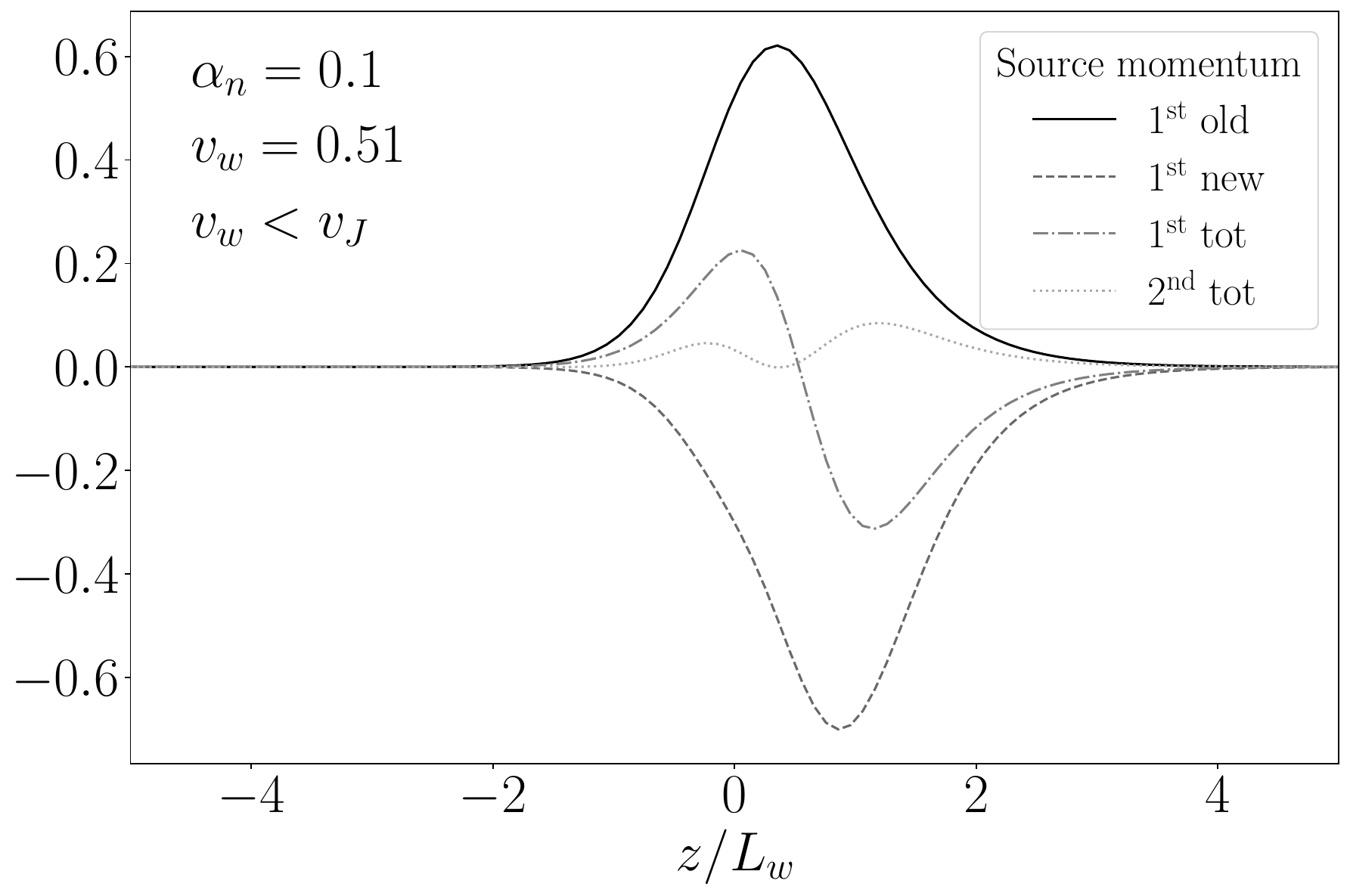}
    \includegraphics[width=.48\textwidth]{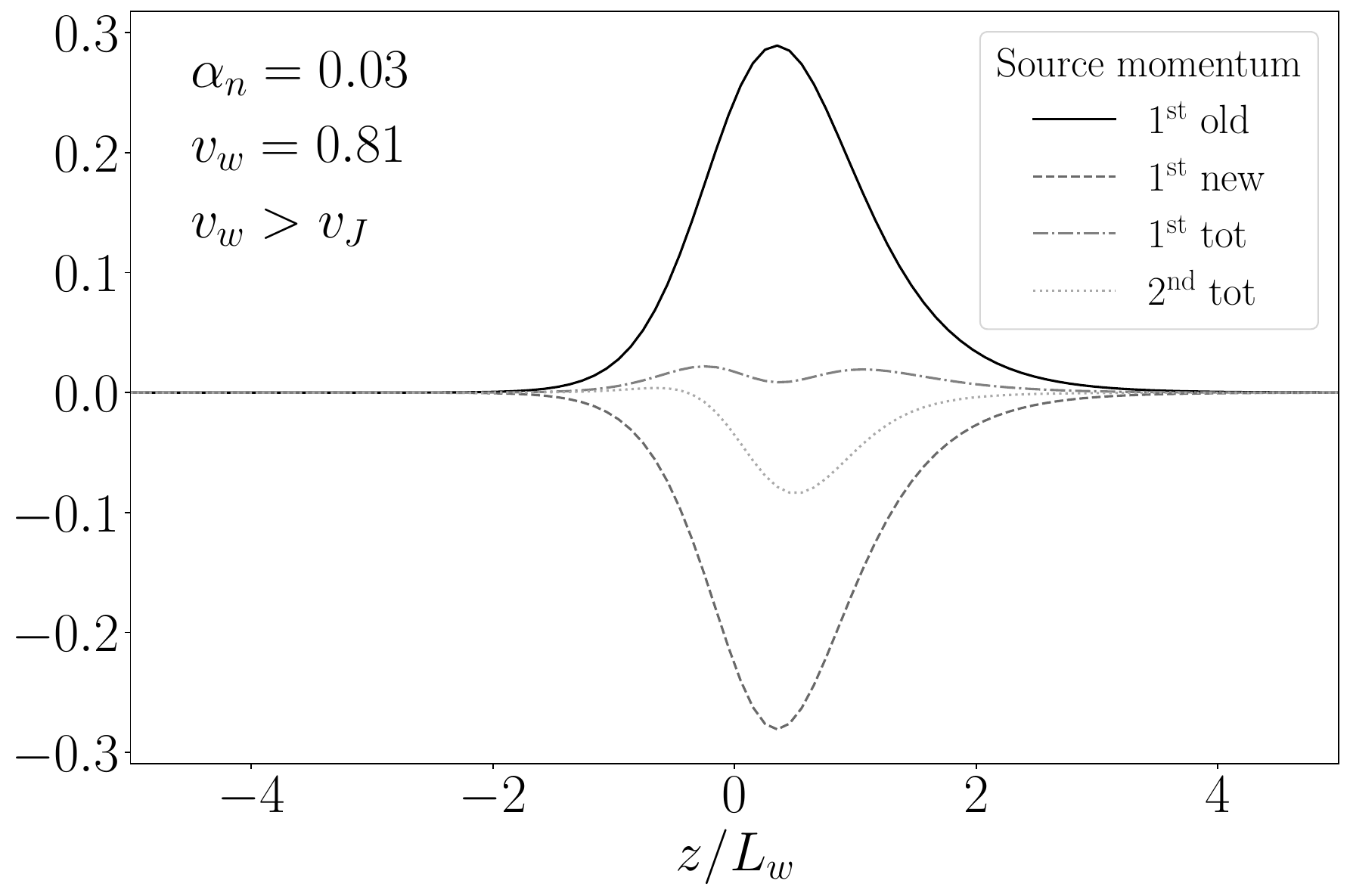}
    \includegraphics[width=.48\textwidth]{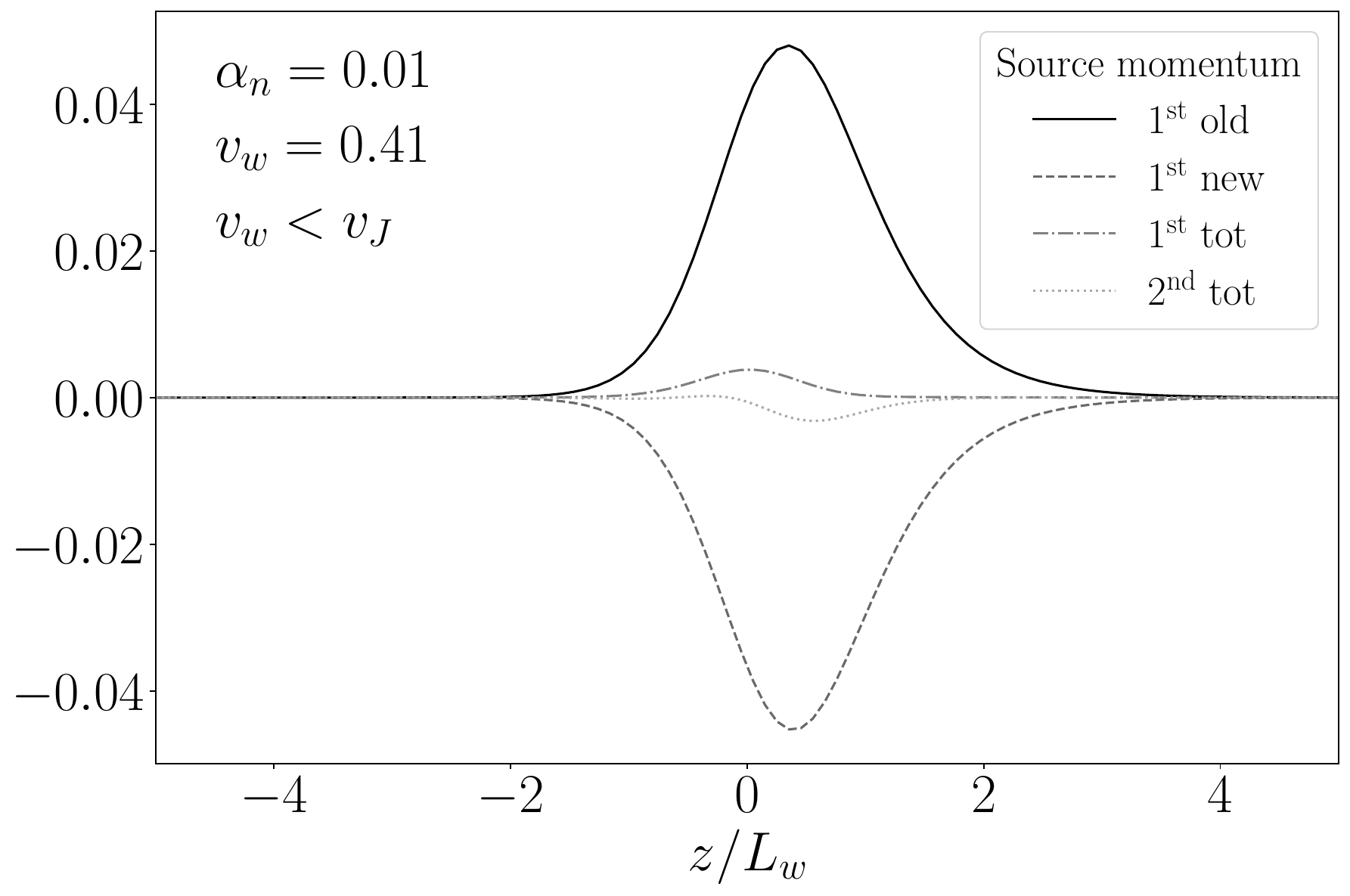}
    \includegraphics[width=.48\textwidth]{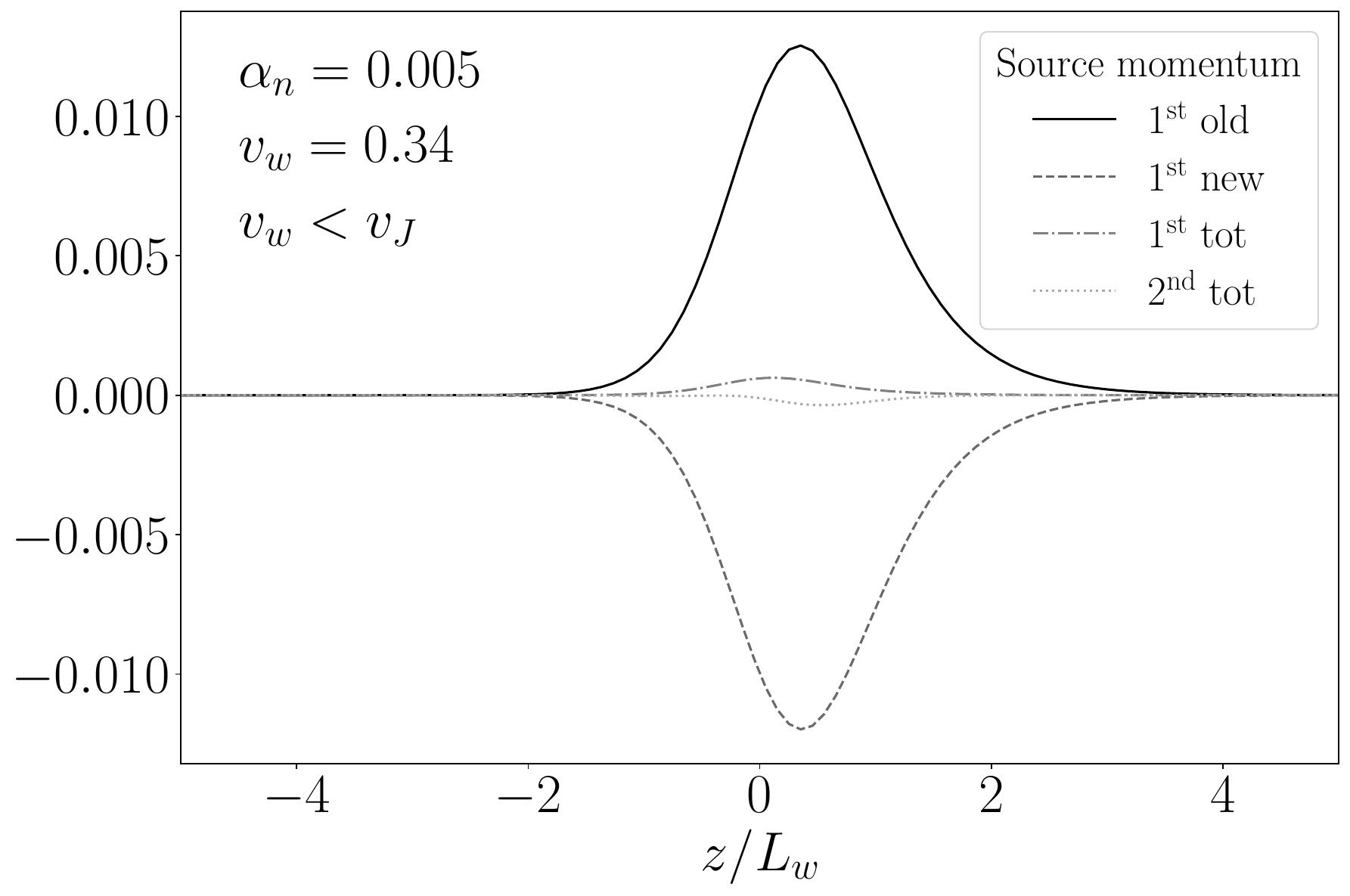}
    \caption{Source momenta of the linearized energy momentum conservation for different solutions at different scales $\Lambda$. The plots are shown for decreasing values of $\alpha_n$. Notice how the source cancellation seems to work better when considering deflagration solutions and also when considering smaller values of $\alpha_n$. 
    }
    \label{fig: source cancellation}
\end{figure}

We discuss now one of the approximation that has been used when solving the linearized Boltzmann system. The two linear combinations in \cref{eq. linearized_en_mom_cons} represents the linearized version of energy-momentum conservation and in the standard procedure are sourced by the mean field term in the potential, thus leading to the singular behavior of the friction at the speed of sound $c_{s,b}$.
As shown in \cref{appendix:: source canc.}, when we consider a $z$-dependent background as defined in \cref{eq:: tot_en-mom_cons} with values of $v_w$ and $L_w$ that solve the Higgs EOM, we find that the two linear combinations on the rhs of \cref{eq. linearized_en_mom_cons} are higher order in the fluctuations. This happens because when the Higgs EOM is satisfied, we can relate the background energy-momentum flux to the distribution functions of the particles in the plasma. When substituting this relation in the explicit form of the sources in \cref{eq. linearized_en_mom_cons} we obtain a leading order cancellation. This is why we enforce $\chi \cdot A \cdot q' =  0$ in the BE solver. However, as we mentioned before, we neglect the mass dependence in the coefficients of the Boltzmann equations and 
also choose a specific {\em Ansatz} for the Higgs field. Both of these approximations explicitly undermine an exact cancellation.

In \cref{fig: source cancellation} we check numerically this approximation. The plots are listed in order of decreasing $\alpha_n$, starting from $\alpha_n \simeq 0.1$. Only the second one represents the case of a detonation, as labeled in the figure by $v_w > v_J$. The continuous, dashed and dashed-dotted grey lines represent respectively the \textit{old}, \textit{new} and \textit{total} contributions to the first linear combination of sources $\mathcal{S}_1$. The dotted line represents the \textit{total} contribution to the second linear combination of sources $\mathcal{S}_2$, and it comes only from the equilibrium backreaction of the fluid.
We see from the plots that the total cancellation depends on the strength of the phase transition, as the two source terms do. 

Focusing on the deflagration cases, we can see that the cancellation is significantly worse for stronger phase transitions (higher $\alpha_n$), while it leaves smaller discrepancies when going to weaker phase transitions. It is important to stress that the condition that ensures the disappearance of the singularity in the friction is not \cref{eq. linearized_en_mom_cons}, but \cref{eq. linear-en-mom}, which is actually a weaker requirement. So, even if the sources  do not cancel precisely, the integrated ones do. 
Looking now into the detonation case in \cref{fig: source cancellation}, we see that the cancellation is again rather incomplete. Differently from what happens with deflagrations, for this family of solutions the sources do not integrate to zero in this case. 

\section{Conclusions} \label{sec. conclusions}

In this work, we studied the friction  and wall velocity in a first-order phase transition 
of a Standard Model with a low cutoff. We paid special attention to the singularity in the pressure for wall velocities close to the speed of sound~\cite{Dorsch_2022, laurent_first_2022}.

The origin of the singularity is an interplay of the fact that the plasma attains different equilibria on both sides of the wall and a zero in the Liouville operator of the total energy-momentum tensor of the fluid once the Boltzmann equations are linearized. A potential solution to the problem was already mentioned in Ref.~\cite{Dorsch_2022} and recently implemented~\cite{laurent_first_2022, Ekstedt:2024fyq}: introducing a space-dependent background fluid allows deviations from equilibrium to die off on both sides of the wall, potentially avoiding the singularity. This is also the route we followed here and we indeed found that the singularity disappears owing to the fact that the equations with the singular Liouville term became sourceless. Since this equation represents the total energy momentum tensor of the fluid, these equations are also undamped and essentially become algebraic relations between the fluctuations of all species in the plasma. This relation can then be used to eliminate the fluctuations of the light species from the system. The problem of the singularity is then resolved in this setup.

Let us now comment in detail on the differences between our approach and the one 
used in~\cite{laurent_first_2022} and \cite{Ekstedt:2024fyq}. One main difference we can identify is that both these works use Chebyshev polynomials for the deviations 
from equilibrium instead of powers of the four-momentum in the flow Ansatz. They assume that the light species (that are not coupled to the wall) remain in equilibrium. Furthermore, the equations are not solved by taking moments but
using a collocation method. The choice of the basis has in principle several consequences. 
In particular, the four-momenta in the fluid Ansatz represent more physical quantities. For example, deviations from equilibrium require $\delta \supset p^\mu \delta u_\mu$ which is not as easily represented in terms of Chebyshev polynomials in $p_\perp$ and $p_{||}$.  Moreover, neglecting the fluctuations in the light species also leads to a 
different behavior. The system does not contain an equation that represents the total energy-momentum of the fluctuations and the singularity will be in any case absent from the system due to this simplification. 

Overall, once the space-dependent background is introduced one would expect that 
the singularity is removed from the system, and this choice of basis, as well as neglecting 
the deviations in the light fields, should be justifiable (see also Sec.~\ref{sec. results and discussion} for the quantitative impact of the light species on the pressure).
On the other hand, \cite{Ekstedt:2024fyq} reported significantly lower wall velocities compared to former results in the literature. At this stage many technical details are implemented differently in the two methods (for example how vanishing fluctuations are enforced at $z \to \pm \infty$). This unfortunately hinders a direct comparison of the intermediate results of these two methods.

\section*{Acknowledgements}

We thank Jorinde van de Vis and Philipp Schicho for discussions.
TK and EP acknowledge
support by the Deutsche Forschungsgemeinschaft (DFG, German Research Foundation) un-
der Germany’s Excellence Strategy – EXC 2121 “Quantum Universe” - 390833306.
This work was financed in part by the Coordena\c{c}\~ao de Aperfei\c{c}oamento de Pessoal de N\'ivel Superior - Brasil (CAPES) - Finance Code 001.
\newpage
\appendix 
\section{New source terms coming from the background} \label{app: sources_en_mom}

\vspace{5mm} \noindent The linearized Boltzmann Equations can also be obtained by the four divergence $\partial_\mu$ of the linearized energy momentum tensor $T^{\mu \nu}$ and particle current $J^{\mu}$, i.e from the flux of particles number and energy-momentum, which can be written as 
\begin{equation} 
    \partial_\mu J^{\mu}_i + \mathcal{C}_i = 0,
\end{equation}
\begin{equation} \label{eq:: energy-momentum flux}
    \partial_\mu T^{\mu \nu}_i + \mathcal{C}_i^\nu = \partial^\nu m_i^2 \int \frac{d^3 p }{(2\pi)^3} \frac{1}{2 E_i} f_i(p^\mu, x^\mu).
\end{equation}
If we focus on the energy-momentum one and linearize it with respect to the fluctuations $\delta \mu$, $\delta T /T$ and $\delta u_{\lambda}$, as done in the Appendix of \cite{Konstandin:2014zta}, we can write
\begin{equation} \label{App:: eq. linear. EM}
\begin{aligned}
T^{\mu \nu} = & T_{\text{bg}}^{\mu \nu}+\bar{T}_{ \text{bg}}^{\mu \nu} \delta \mu+\beta \bar{M}_{ \text{bg}}^{\mu \nu \lambda}\left(\delta \tau u_\lambda+\delta u_\lambda\right) \\
= & T_{\text{bg}}^{\mu \nu}-T \partial_T T_{\text{bg}}^{\mu \nu} \delta \tau \\
& +\left(g^{\mu \nu} n_{\text{bg}} -u^\mu u^\nu\left(n_{\text{bg}} +T \partial_T n_{\text{bg}} \right)\right) \delta \mu \\
& -\left(\delta u^\mu u^\nu+u^\mu \delta u^\nu\right) \omega_{\text{bg}} 
\end{aligned}
\end{equation}
where the barred tensors are the defined as particular kind of average quantities over the particles distribution functions
\begin{equation}
    \langle O \rangle \equiv \int \frac{d^3 p}{(2 \pi)^3} \frac{1}{E_i} \, O \, f_i^{\text{eq}} \hspace{5mm} 
    \left [ O \right ] \equiv \int \frac{d^3 p}{(2 \pi)^3} \frac{1}{E_i} \, O \, (f_i^{\text{eq}})' \, ,
\end{equation}
with $\Bar{J}^\mu \equiv \left [ p^\mu \right ] $, $\Bar{T}^{\mu \nu} \equiv \left [ p^\mu p^\nu \right ]$ and $\Bar{M}^{\mu \nu \lambda} \equiv \left [ p^\mu p^\nu p^\lambda \right ]$ and 
\begin{equation}
    n_{\text{bg}} = \frac{\zeta (3) T^3}{4\pi^2}\times\left\{\begin{array}{ll}
        4, & \text{for bosons}\\
        3,& \text{for fermions.}\end{array}\right.
\end{equation}
The source term comes from the derivative acting on $T^{\mu \nu}_{\text{bg}}$. The left hand side of \cref{eq:: energy-momentum flux} gives 
\begin{equation}
\begin{aligned}
\partial_\mu T^{\mu \nu}_{\text{bg}} &= \partial_\mu m^2(z) (u^{\mu} u^{\nu} \partial_{m^2} \omega  - g^{\mu \nu} \partial_{m^2} P) \\
& +\partial_\mu T(z) ( u^{\mu} u^{\nu} \partial_T \omega  - g^{\mu \nu} \partial_T P) + \omega (\partial_\mu u^{\mu}(z)) u^{\nu} + \omega u^{\mu}(\partial_\mu u^{\nu}(z)).
\end{aligned}
\end{equation}
When projecting this along $u_\nu$, we obtain
\begin{equation} \label{eq:: 1st mom sources equilibrium}
\begin{aligned}
 u_\nu \partial_\mu T^{\mu \nu}_{\text{bg}} &\simeq u^\mu \partial_\mu m^2(z) (\partial_{m^2} \omega - \partial_{m^2} P) + u^\mu \partial_\mu T(z) (\partial_T \omega - \partial_T P) + \\
 &+\omega (\partial_\mu u^{\mu}(z)) + \omega u_\nu u^{\mu}(\partial_\mu u^{\nu}(z)).
\end{aligned}
\end{equation}
Considering the definitions of $P_{\text{bg}}$, $\omega_{\text{bg}}$ and the identity $g_{\mu \nu} T^{\mu \nu} = m^2 N$, where $N = \langle 1 \rangle$, 
we can see that in this case the $\propto \partial_z m^2(z)$ term on the lhs sums up with the term on the rhs in \cref{eq:: energy-momentum flux} to give the original source term, since the relation $\partial_{m^2} P = -\frac{1}{2} N$ holds if $g_{\mu \nu} T^{\mu \nu} = m^2 N$. The other terms are $\propto \partial_z T(z), \, \partial_z v(z)$ and represent the non-trivial spatial behavior of the background fluid. We can define
\begin{equation}
\gamma_w \frac{4}{3} a T^4 \biggr(3 v_w \beta \partial_z T(z) + \gamma_w^2 \partial_z v(z) \biggr) \equiv - \sum_{i} S_i^{\text{new}},
\end{equation}
so that the first equation for the energy-momentum tensor will be sourced by $\sum_i S_i^{\text{old}} + \sum_i S_i^{\text{new}} \equiv \mathcal{S}_1$. Projecting then along $\Bar{u}_{\nu}$, one obtains 
\begin{equation} 
\begin{aligned}
\Bar{u}_{\nu} \partial_\mu T^{\mu \nu}_{\text{bg}} &\simeq 
- \gamma_w \partial_z m^2(z) \partial_{m^2} P - \gamma_w \partial_z T(z) \partial_T P - \omega(z) \gamma_w^3 v_w \partial_z v(z) .
\end{aligned} 
\end{equation}
In this case, due to the opposite sign in front of the first term with respect to \cref{eq:: 1st mom sources equilibrium}, the two pieces proportional to $\partial_z m^2(z)$ cancel out, leaving
\begin{equation} \label{eq. new_sources_TOT2}
\mathcal{S}_2 \equiv - \gamma_w \frac{4}{3} a T^4 \biggr ( \beta \partial_z T(z) + \gamma_w^2 v_w \partial_z v(z) \biggr ).
\end{equation}

The functions $\mathcal{S}_1$ and $\mathcal{S}_2$ are the total sources of the linearized Boltzmann equation that represents the energy-momentum conservation of the fluid, since they take into account all the degrees of freedom present in the plasma. These functions can be used, since the functional dependence turns out to be the same, even for the single particles linearized equations in \cref{eq. Boltz. eq heavy guys OLD}, after taking care of the numerical prefactors coming from the different (bosonic or fermionic) degrees of freedom, as done in \cref{appendix:: linearized BEs}. As we are going to show in \cref{appendix:: source canc.}, these functions vanishes at linear order ones the Higgs EoM is solved.

\section{Explicit on-shell source cancellation} \label{appendix:: source canc.}

Here we want to look at the behavior of the source terms $\mathcal{S}_1$ and $\mathcal{S}_2$ when taken on-shell. These can be extracted from the energy-momentum flux in \cref{eq:: energy-momentum flux} if we consider the parts that do not involve the out of equilibrium fluctuations $\delta f_i$, namely
\begin{equation} \label{eq:: sources_from Kad-Baym}
    \partial^\nu m_i^2 \int \frac{d^3 p }{(2\pi)^3} \frac{1}{2 E_i} f_i^{\text{bg}}(p^\mu, x^\mu) - \partial_\mu T^{\mu \nu}_{i,\text{bg}}.
\end{equation}
We defined our \textit{background} through its energy-momentum tensor conservation in \cref{eq:: tot_en-mom_cons}, which involves only equilibrium contributions. This is equivalent to solving 
\begin{equation} \label{eq:: background_def}
    \sum_i \partial_\mu T^{\mu \nu}_{i,\text{bg}} + \partial_\mu T^{\mu \nu}_\phi = 0, 
\end{equation}
with the \textit{background} energy-momentum tensor as defined in \cref{eq. equil.energy-mom tensor} and the Higgs one as defined in \cref{eq:: higgs en_mom_tensor}. The collision term drops out when the sum over all the particles is taken, i.e $\sum_i \mathcal{C}^\nu_i = 0$. The above equation is solved to find the temperature and velocity profiles of this background field, namely $T_{bg}(z)$ and $v_{bg}(z)$. When solving for $v_w$ and $L_w$, we have to actually tackle the whole system constituted by \cref{eq:: tot_en-mom_cons}, \cref{eq. Boltz. eq heavy guys OLD} and the Higgs EoM, \cref{eq: KG eq_Higgs}. When this last one is satisfied, then 
\begin{equation}
    \partial^\nu \phi \left( \Box \phi + \frac{d V_0}{d \phi}\right) = - \partial^\nu \phi \sum_i \frac{d m_i^2}{d \phi} \int \frac{d^3 p}{(2 \pi)^3} \frac{1}{2 E_i} f_i(p^\mu, x^\mu).
\end{equation} 
If we now consider how the \textit{background} is defined in \cref{eq:: tot_en-mom_cons}, or equivalently in \cref{eq:: background_def}, and we put everything back in \eqref{eq:: sources_from Kad-Baym}, we find
\begin{equation}
    \begin{aligned}
        & \sum_i \partial^\nu m_i^2 \int \frac{d^3 p }{(2\pi)^3} \frac{1}{2 E_i} \left ( f_i^{\text{bg}}(p^\mu, x^\mu) - f_i(p^\mu, x^\mu) \right ) =\\
        &\sum_i \partial^\nu m_i^2 \int \frac{d^3 p }{(2\pi)^3} \frac{1}{2 E_i} \delta f_i(p^\mu, x^\mu) \simeq 0,
    \end{aligned}
\end{equation}
where the second equality follows from the fact that the second line is higher order in the fluctuations, i.e $\mathcal{O}(\partial m^2 \cdot \delta f)$. We just found that, at linear order in perturbation theory, the linear combinations of the BEs that represent energy-momentum conservation, i.e \cref{eq. linearized_en_mom_cons}, are unsourced. In other words, the new terms compensate the forces in the equations of total energy-momentum conservation and impose that the fluctuations of the light species vanish away from the phase transition front.

\section{Solving the linearized Boltzmann system} \label{appendix:: linearized BEs}

After linearizing the fluctuations of the distribution function $ f $, one can express the system of equations in the wall frame as follows:

\begin{equation}
\begin{aligned}
\label{eq:BEall}
    A_t \left( q_t^{\prime} + q_{\mathrm{light}}^{\prime} \right) + \Gamma_t q_t &= \tilde{S}_t, \\
    A_W \left( q_W^{\prime} + q_{\mathrm{light}}^{\prime} \right) + \Gamma_W q_W &= \tilde{S}_W, \\
    A_{\mathrm{bg}} q_{\mathrm{light}}^{\prime} + \Gamma_{\mathrm{light}, t} q_t + \Gamma_{\mathrm{light}, W} q_W &= 0.
\end{aligned}
\end{equation}
Assuming $ \mu_{\mathrm{light}} = 0 $, one of these equations becomes redundant. 
For obvious reasons, we keep the two equations representing the energy-momentum 
tensor of the light species and discard the current conservation of the light species. 
The damping terms of the light degrees of freedom then relate to the ones from the top and $W$
as explained in the main text.

The remaining eight equations, applied to the tops, Ws, and light sectors, can be summed over to eliminate the collision terms as well as the total source term. This yields
\begin{equation}
(\bar A_t + \bar A_W + \bar A_{\mathrm{light}} ) \tilde q_{\mathrm{light}}^\prime 
+ \tilde A_t  q_t^{\prime} + \tilde A_W  q_W^{\prime}  = 0 \, ,
\end{equation}
where we introduced the notation that $\bar X$ denotes the $2\times2$ submatrix
that neglects the row and column corresponding to the chemical potential. Likewise, $\tilde X$ is the $2\times3$ submatrix that only neglects the row corresponding to the chemical potential. This relation can be inverted and the light degrees of freedom can be then eliminated from the first two equations of (\ref{eq:BEall}). We will call $\mathcal{A}$ the matrix that multiplies $q^\prime$ after $q^\prime_\text{light}$ has been eliminated from eq.~\eqref{eq:BEall}.

The gamma matrices in equation (\ref{eq:BEall}) are derived from the first-order moments of the collision terms \cite{dorsch_bubble_2023}, which are limited to $2\to 2$ processes, each characterized by an amplitude $\mathcal{M}_{pk\to p^\prime k^\prime}$. The collision term is expressed as
\begin{equation}
\begin{split}
    \mathcal{C}[f_p] = \frac{1}{2}\sum_{\rm processes} &\int
				\dfrac{d^3k\, d^3p^\prime d^3k^\prime}{(2\pi)^{9} 2E_k\, 2E_{p^\prime}\, 2E_{k^\prime}} |\mathcal{M}_{pk\to p^\prime k^\prime}|^2 (2\pi)^4 \delta^4(p\!+\! k\!-\!p^\prime\!-\!k^\prime) \mathcal{P}_{pk \to p^\prime k^\prime},
	\label{eq:full_coll}
\end{split}
\end{equation}
where
\begin{equation}
	\mathcal{P}_{pk \to p^\prime k^\prime} \equiv f_p f_k (1\pm f_{p^\prime}) (1\pm f_{k^\prime}) -
							f_{p^\prime} f_{k^\prime} (1\pm f_p) (1\pm f_k)
	\label{eq:P_occupation}
\end{equation}
is the population factor, accounting for how the number density of reactants and products in the plasma influences the reaction rate.

To solve the system in equation (\ref{eq:BEall}), one can apply the standard Green's method \cite{Cline:2020jre,Lewicki:2021pgr,Dorsch_2022}, integrating now over the $z$-dependent components of the source terms
\begin{equation}
    S(z) = \gamma_w v_w\frac{m^\prime m}{2\pi^2 T^2}
    \begin{pmatrix}
        c_1 \\
        c_2 \\
        0
    \end{pmatrix}
    - \dfrac{1}{ 3\cdot 2 \pi^2} \gamma_w
    \begin{pmatrix}
    c_3 \left(3v_w \dfrac{T'}{T} + \gamma_w^2 v'\right) \\
    c_4 \left(3v_w \dfrac{T'}{T} + \gamma_w^2 v'\right) \\
    c_4 \left(\dfrac{T'}{T} + \gamma^2_w v_w v^\prime\right)
    \end{pmatrix},
\end{equation}
where the first term is the old source and the last one is the new contribution as discussed in section~\ref{sec. new_way_Boltz}. 
This then leads to the fluctuations
\begin{equation}
    \begin{aligned}
    q(z) =\, &\chi\cdot \int_{-\infty}^\infty 
				e^{-\lambda  (z-z^\prime)}
				\cdot \theta(\lambda(z-z^\prime))\cdot \text{sign}(\lambda)\cdot \left(\chi^{-1}\cdot \mathcal{A}^{-1}\cdot S(z)\right) dz^\prime
    \end{aligned}
\end{equation}
where $\chi$ is a matrix whose columns are the eigenvectors of the matrix $\mathcal{A}^{-1}\cdot \Gamma/L_w $, and $\lambda$ a diagonal matrix with the corresponding eigenvalues. Note that the source terms and also the matrix $\mathcal{A}$ are different for bosons and fermions due to the difference in the coefficients $c_n$, see eq.~\eqref{eq:p_ints}.

\newpage

\providecommand{\href}[2]{#2}\begingroup\raggedright\endgroup

\begin{thebibliography}{10}
	
	\bibitem{LIGOScientific:2016aoc}
	{\scshape LIGO Scientific, Virgo} collaboration, \emph{{Observation of
			Gravitational Waves from a Binary Black Hole Merger}},
	\href{https://doi.org/10.1103/PhysRevLett.116.061102}{\emph{Phys. Rev. Lett.}
		{\bfseries 116} (2016) 061102}
	[\href{https://arxiv.org/abs/1602.03837}{{\ttfamily 1602.03837}}].
	
	\bibitem{LIGOScientific:2018mvr}
	{\scshape LIGO Scientific, Virgo} collaboration, \emph{{GWTC-1: A
			Gravitational-Wave Transient Catalog of Compact Binary Mergers Observed by
			LIGO and Virgo during the First and Second Observing Runs}},
	\href{https://doi.org/10.1103/PhysRevX.9.031040}{\emph{Phys. Rev. X}
		{\bfseries 9} (2019) 031040}
	[\href{https://arxiv.org/abs/1811.12907}{{\ttfamily 1811.12907}}].
	
	\bibitem{LIGOScientific:2021usb}
	{\scshape LIGO Scientific, VIRGO} collaboration, \emph{{GWTC-2.1: Deep Extended
			Catalog of Compact Binary Coalescences Observed by LIGO and Virgo During the
			First Half of the Third Observing Run}},
	\href{https://arxiv.org/abs/2108.01045}{{\ttfamily 2108.01045}}.
	
	\bibitem{LIGOScientific:2021djp}
	{\scshape LIGO Scientific, VIRGO, KAGRA} collaboration, \emph{{GWTC-3: Compact
			Binary Coalescences Observed by LIGO and Virgo During the Second Part of the
			Third Observing Run}},  \href{https://arxiv.org/abs/2111.03606}{{\ttfamily
			2111.03606}}.
	
	\bibitem{NANOGrav:2023gor}
	{\scshape NANOGrav} collaboration, \emph{{The NANOGrav 15 yr Data Set: Evidence
			for a Gravitational-wave Background}},
	\href{https://doi.org/10.3847/2041-8213/acdac6}{\emph{Astrophys. J. Lett.}
		{\bfseries 951} (2023) L8}
	[\href{https://arxiv.org/abs/2306.16213}{{\ttfamily 2306.16213}}].
	
	\bibitem{EPTA:2023fyk}
	{\scshape EPTA, InPTA:} collaboration, \emph{{The second data release from the
			European Pulsar Timing Array - III. Search for gravitational wave signals}},
	\href{https://doi.org/10.1051/0004-6361/202346844}{\emph{Astron. Astrophys.}
		{\bfseries 678} (2023) A50}
	[\href{https://arxiv.org/abs/2306.16214}{{\ttfamily 2306.16214}}].
	
	\bibitem{Caprini_2020}
	C.~Caprini, M.~Chala, G.~C. Dorsch, M.~Hindmarsh, S.~J. Huber, T.~Konstandin
	et~al., \emph{Detecting gravitational waves from cosmological phase
		transitions with lisa: an update},
	\href{https://doi.org/10.1088/1475-7516/2020/03/024}{\emph{Journal of
			Cosmology and Astroparticle Physics} {\bfseries 2020} (2020) 024–024}.
	
	\bibitem{Arcadi_2024}
	G.~Arcadi, G.~C. Dorsch, J.~P. Neto, F.~S. Queiroz and Y.~Oviedo-Torres,
	\emph{Probing a dark sector with collider physics, direct detection, and
		gravitational waves},
	\href{https://doi.org/10.1016/j.physletb.2023.138382}{\emph{Physics Letters
			B} {\bfseries 848} (2024) 138382}.
	
	\bibitem{Bodeker:2009qy}
	D.~Bodeker and G.~D. Moore, \emph{{Can electroweak bubble walls run away?}},
	\href{https://doi.org/10.1088/1475-7516/2009/05/009}{\emph{JCAP} {\bfseries
			05} (2009) 009} [\href{https://arxiv.org/abs/0903.4099}{{\ttfamily
			0903.4099}}].
	
	\bibitem{Hoche:2020ysm}
	S.~H\"oche, J.~Kozaczuk, A.~J. Long, J.~Turner and Y.~Wang, \emph{{Towards an
			all-orders calculation of the electroweak bubble wall velocity}},
	\href{https://doi.org/10.1088/1475-7516/2021/03/009}{\emph{JCAP} {\bfseries
			03} (2021) 009} [\href{https://arxiv.org/abs/2007.10343}{{\ttfamily
			2007.10343}}].
	
	\bibitem{Azatov:2020ufh}
	A.~Azatov and M.~Vanvlasselaer, \emph{{Bubble wall velocity: heavy physics
			effects}}, \href{https://doi.org/10.1088/1475-7516/2021/01/058}{\emph{JCAP}
		{\bfseries 01} (2021) 058}
	[\href{https://arxiv.org/abs/2010.02590}{{\ttfamily 2010.02590}}].
	
	\bibitem{Ai:2024shx}
	W.-Y. Ai, X.~Nagels and M.~Vanvlasselaer, \emph{{Criterion for ultra-fast
			bubble walls: the impact of hydrodynamic obstruction}},
	\href{https://doi.org/10.1088/1475-7516/2024/03/037}{\emph{JCAP} {\bfseries
			03} (2024) 037} [\href{https://arxiv.org/abs/2401.05911}{{\ttfamily
			2401.05911}}].
	
	\bibitem{Long:2024sqg}
	A.~J. Long and J.~Turner, \emph{{Thermal pressure on ultrarelativistic bubbles
			from a semiclassical formalism}},
	\href{https://arxiv.org/abs/2407.18196}{{\ttfamily 2407.18196}}.
	
	\bibitem{DeCurtis:2022hlx}
	S.~De~Curtis, L.~D. Rose, A.~Guiggiani, A.~G. Muyor and G.~Panico,
	\emph{{Bubble wall dynamics at the electroweak phase transition}},
	\href{https://doi.org/10.1007/JHEP03(2022)163}{\emph{JHEP} {\bfseries 03}
		(2022) 163} [\href{https://arxiv.org/abs/2201.08220}{{\ttfamily
			2201.08220}}].
	
	\bibitem{DeCurtis:2023hil}
	S.~De~Curtis, L.~Delle~Rose, A.~Guiggiani, A.~Gil~Muyor and G.~Panico,
	\emph{{Collision integrals for cosmological phase transitions}},
	\href{https://doi.org/10.1007/JHEP05(2023)194}{\emph{JHEP} {\bfseries 05}
		(2023) 194} [\href{https://arxiv.org/abs/2303.05846}{{\ttfamily
			2303.05846}}].
	
	\bibitem{DeCurtis:2024hvh}
	S.~De~Curtis, L.~Delle~Rose, A.~Guiggiani, A.~Gil~Muyor and G.~Panico,
	\emph{{Non-linearities in cosmological bubble wall dynamics}},
	\href{https://doi.org/10.1007/JHEP05(2024)009}{\emph{JHEP} {\bfseries 05}
		(2024) 009} [\href{https://arxiv.org/abs/2401.13522}{{\ttfamily
			2401.13522}}].
	
	\bibitem{Moore:1995si}
	G.~D. Moore and T.~Prokopec, \emph{{How fast can the wall move? A Study of the
			electroweak phase transition dynamics}},
	\href{https://doi.org/10.1103/PhysRevD.52.7182}{\emph{Phys. Rev. D}
		{\bfseries 52} (1995) 7182}
	[\href{https://arxiv.org/abs/hep-ph/9506475}{{\ttfamily hep-ph/9506475}}].
	
	\bibitem{Konstandin:2014zta}
	T.~Konstandin, G.~Nardini and I.~Rues, \emph{{From Boltzmann equations to
			steady wall velocities}},
	\href{https://doi.org/10.1088/1475-7516/2014/09/028}{\emph{JCAP} {\bfseries
			09} (2014) 028} [\href{https://arxiv.org/abs/1407.3132}{{\ttfamily
			1407.3132}}].
	
	\bibitem{Dorsch:2021ubz}
	G.~C. Dorsch, S.~J. Huber and T.~Konstandin, \emph{{On the wall velocity
			dependence of electroweak baryogenesis}},
	\href{https://doi.org/10.1088/1475-7516/2021/08/020}{\emph{JCAP} {\bfseries
			08} (2021) 020} [\href{https://arxiv.org/abs/2106.06547}{{\ttfamily
			2106.06547}}].
	
	\bibitem{Dorsch_2022}
	G.~C. Dorsch, S.~J. Huber and T.~Konstandin, \emph{A sonic boom in bubble wall
		friction}, \href{https://doi.org/10.1088/1475-7516/2022/04/010}{\emph{Journal
			of Cosmology and Astroparticle Physics} {\bfseries 2022} (2022) 010}.
	
	\bibitem{dorsch_bubble_2023}
	G.~C. Dorsch and D.~A. Pinto, \emph{Bubble wall velocities with an extended
		fluid {Ansatz}},  Dec., 2023.
	
	\bibitem{DeGroot:1980dk}
	S.~R. De~Groot, W.~A. Van~Leeuwen and C.~G. Van~Weert, \emph{{Relativistic
			Kinetic Theory. Principles and Applications}}. North-Holland Pub. Co. ; Sole
	distributors for the USA and Canada, Elsevier North-Holland., 1980.
	
	\bibitem{laurent_first_2022}
	B.~Laurent and J.~M. Cline, \emph{First principles determination of bubble wall
		velocity}, \href{https://doi.org/10.1103/PhysRevD.106.023501}{\emph{Physical
			Review D} {\bfseries 106} (2022) 023501}.
	
	\bibitem{Konstandin:2010dm}
	T.~Konstandin and J.~M. No, \emph{{Hydrodynamic obstruction to bubble
			expansion}}, \href{https://doi.org/10.1088/1475-7516/2011/02/008}{\emph{JCAP}
		{\bfseries 02} (2011) 008} [\href{https://arxiv.org/abs/1011.3735}{{\ttfamily
			1011.3735}}].
	
	\bibitem{Ai:2021kak}
	W.-Y. Ai, B.~Garbrecht and C.~Tamarit, \emph{{Bubble wall velocities in local
			equilibrium}},
	\href{https://doi.org/10.1088/1475-7516/2022/03/015}{\emph{JCAP} {\bfseries
			03} (2022) 015} [\href{https://arxiv.org/abs/2109.13710}{{\ttfamily
			2109.13710}}].
	
	\bibitem{Ai:2023see}
	W.-Y. Ai, B.~Laurent and J.~van~de Vis, \emph{{Model-independent bubble wall
			velocities in local thermal equilibrium}},
	\href{https://doi.org/10.1088/1475-7516/2023/07/002}{\emph{JCAP} {\bfseries
			07} (2023) 002} [\href{https://arxiv.org/abs/2303.10171}{{\ttfamily
			2303.10171}}].
	
	\bibitem{Lewicki:2021pgr}
	M.~Lewicki, M.~Merchand and M.~Zych, \emph{{Electroweak bubble wall expansion:
			gravitational waves and baryogenesis in Standard Model-like thermal plasma}},
	\href{https://doi.org/10.1007/JHEP02(2022)017}{\emph{JHEP} {\bfseries 02}
		(2022) 017} [\href{https://arxiv.org/abs/2111.02393}{{\ttfamily
			2111.02393}}].
	
	\bibitem{Cline:2021iff}
	J.~M. Cline, A.~Friedlander, D.-M. He, K.~Kainulainen, B.~Laurent and
	D.~Tucker-Smith, \emph{{Baryogenesis and gravity waves from a UV-completed
			electroweak phase transition}},
	\href{https://doi.org/10.1103/PhysRevD.103.123529}{\emph{Phys. Rev. D}
		{\bfseries 103} (2021) 123529}
	[\href{https://arxiv.org/abs/2102.12490}{{\ttfamily 2102.12490}}].
	
	\bibitem{Konstandin:2013caa}
	T.~Konstandin, \emph{{Quantum Transport and Electroweak Baryogenesis}},
	\href{https://doi.org/10.3367/UFNe.0183.201308a.0785}{\emph{Phys. Usp.}
		{\bfseries 56} (2013) 747} [\href{https://arxiv.org/abs/1302.6713}{{\ttfamily
			1302.6713}}].
	
	\bibitem{Espinosa:2010hh}
	J.~R. Espinosa, T.~Konstandin, J.~M. No and G.~Servant, \emph{{Energy Budget of
			Cosmological First-order Phase Transitions}},
	\href{https://doi.org/10.1088/1475-7516/2010/06/028}{\emph{JCAP} {\bfseries
			06} (2010) 028} [\href{https://arxiv.org/abs/1004.4187}{{\ttfamily
			1004.4187}}].
	
	\bibitem{Hindmarsh:2020hop}
	M.~B. Hindmarsh, M.~L\"uben, J.~Lumma and M.~Pauly, \emph{{Phase transitions in
			the early universe}},
	\href{https://doi.org/10.21468/SciPostPhysLectNotes.24}{\emph{SciPost Phys.
			Lect. Notes} {\bfseries 24} (2021) 1}
	[\href{https://arxiv.org/abs/2008.09136}{{\ttfamily 2008.09136}}].
	
	\bibitem{Coleman:1977py}
	S.~R. Coleman, \emph{{The Fate of the False Vacuum. 1. Semiclassical Theory}},
	\href{https://doi.org/10.1103/PhysRevD.16.1248}{\emph{Phys. Rev. D}
		{\bfseries 15} (1977) 2929}.
	
	\bibitem{Linde:1981zj}
	A.~D. Linde, \emph{{Decay of the False Vacuum at Finite Temperature}},
	\href{https://doi.org/10.1016/0550-3213(83)90072-X}{\emph{Nucl. Phys. B}
		{\bfseries 216} (1983) 421}.
	
	\bibitem{PhysRevD.25.2074}
	P.~J. Steinhardt, \emph{Relativistic detonation waves and bubble growth in
		false vacuum decay},
	\href{https://doi.org/10.1103/PhysRevD.25.2074}{\emph{Phys. Rev. D}
		{\bfseries 25} (1982) 2074}.
	
	\bibitem{Giese_2020}
	F.~Giese, T.~Konstandin and J.~v. de~Vis, \emph{Model-independent energy budget
		of cosmological first-order phase transitions—a sound argument to go beyond
		the bag model},
	\href{https://doi.org/10.1088/1475-7516/2020/07/057}{\emph{Journal of
			Cosmology and Astroparticle Physics} {\bfseries 2020} (2020) 057–057}.
	
	\bibitem{Leitao_2011}
	L.~Leitao and A.~Mégevand, \emph{Spherical and non-spherical bubbles in
		cosmological phase transitions},
	\href{https://doi.org/10.1016/j.nuclphysb.2010.11.012}{\emph{Nuclear Physics
			B} {\bfseries 844} (2011) 450–470}.
	
	\bibitem{Balaji:2020yrx}
	S.~Balaji, M.~Spannowsky and C.~Tamarit, \emph{{Cosmological bubble friction in
			local equilibrium}},
	\href{https://doi.org/10.1088/1475-7516/2021/03/051}{\emph{JCAP} {\bfseries
			03} (2021) 051} [\href{https://arxiv.org/abs/2010.08013}{{\ttfamily
			2010.08013}}].
	
	\bibitem{Grojean:2004xa}
	C.~Grojean, G.~Servant and J.~D. Wells, \emph{{First-order electroweak phase
			transition in the standard model with a low cutoff}},
	\href{https://doi.org/10.1103/PhysRevD.71.036001}{\emph{Phys. Rev. D}
		{\bfseries 71} (2005) 036001}
	[\href{https://arxiv.org/abs/hep-ph/0407019}{{\ttfamily hep-ph/0407019}}].
	
	\bibitem{Cline:2020jre}
	J.~M. Cline and K.~Kainulainen, \emph{{Electroweak baryogenesis at high bubble
			wall velocities}},
	\href{https://doi.org/10.1103/PhysRevD.101.063525}{\emph{Phys. Rev. D}
		{\bfseries 101} (2020) 063525}
	[\href{https://arxiv.org/abs/2001.00568}{{\ttfamily 2001.00568}}].
	
	\bibitem{Ekstedt:2024fyq}
	A.~Ekstedt, O.~Gould, J.~Hirvonen, B.~Laurent, L.~Niemi, P.~Schicho et~al.,
	\emph{{How fast does the WallGo? A package for computing wall velocities in
			first-order phase transitions}},
	\href{https://arxiv.org/abs/2411.04970}{{\ttfamily 2411.04970}}.
	
\end{thebibliography}
\end{document}